\documentclass[12pt]{iopart}
\pdfoutput=1
\usepackage{amsfonts}
\usepackage{graphicx}

\usepackage{color}
\usepackage{subfigure}
\usepackage{textcomp}
\usepackage{cite}
\usepackage{float}
\usepackage{soul}

\definecolor{dgreen}{rgb}{0,0.7,0}

\expandafter\let\csname equation*\endcsname\relax
\expandafter\let\csname endequation*\endcsname\relax
\usepackage{amsmath,amssymb}

\usepackage{esint}

\begin{document}

\title[]{Steady-state, relaxation and first-passage properties of a run-and-tumble particle in one-dimension}

\author{Kanaya Malakar$^1$, V. Jemseena$^2$, Anupam Kundu$^2$, K. Vijay Kumar$^2$, Sanjib Sabhapandit$^3$, Satya N. Majumdar$^4$, S. Redner$^5$, Abhishek Dhar$^2$}

\address{$^1$Presidency University, 86/1, College Street, Kolkata 700073, India}
\address{$^2$International Centre for Theoretical Sciences, Tata Institute of Fundamental Research, Bengaluru 560089, India}
\address{$^3$Raman Research Institute, Bangalore 560080, India}
\address{$^4$LPTMS, CNRS, Univ. Paris-Sud, Universit\'{e} Paris-Saclay, 91405 Orsay, France}
\address{$^5$Santa Fe Institute, 1399 Hyde Park Road, Santa Fe, New Mexico 87501, USA}

\begin{abstract}
  We investigate the motion of a run-and-tumble particle (RTP) in one
  dimension.  We find the exact probability distribution of the particle with
  and without diffusion on the infinite line, as well as in a finite
  interval.  In the infinite domain, this probability distribution approaches
  a Gaussian form in the long-time limit, as in the case of a regular
  Brownian particle.  At intermediate times, this distribution exhibits
  unexpected multi-modal forms.  In a finite domain, the probability
  distribution reaches a steady-state form with peaks at the boundaries, in
  contrast to a Brownian particle.  We also study the relaxation to the
  steady-state analytically.  Finally we compute the survival probability of
  the RTP in a semi-infinite domain with an absorbing boundary condition at the origin.  In the finite interval, we compute the
  exit probability and the associated exit times.  We provide numerical
  verification of our analytical results.
\end{abstract}

\section{Introduction}
Active particles are self-driven systems, where the dynamics has a
dissipative and a stochastic part. Their dynamics violates
fluctuation-dissipation relation.  This system naturally breaks detailed
balance and has been widely used to understand various non-equilibrium
phenomena which are driven at the level of individual constituents, for example
motion of bacteria, flocking of birds and vibrated granular
matter~\cite{ramaswamy_active_2017,prost_active_2015,marchetti_hydrodynamics_2013,BB72,BP77,DZ88}. Run-and-tumble
particles (RTPs) and active Brownian particles (ABPs) are the simplest
examples of such active particles and are known to exhibit interesting
features such as non-Boltzmann distributions in the
steady-state~\cite{bechinger_active_2016, cates_when_2013, cates_when_2009,Enculescu_when_2011,
Lee_when_2011, Fily_when_2014, Szamel_when_2014, Solon_when_2015,  Vachier_when_2017, Hermann_when_2017,
romanczuk_active_2012}, clustering \cite{evans2016,evans2017}, spontaneous
segregation of mixtures of active and passive
particles~\cite{stenhammar_activity-induced_2015}, ratchet
effects~\cite{reichhardt_ratchet_2017} and motility-induced phase separation
\cite{cates_motility-induced_2015,PhysRevLett.110.055701,stenhammar_continuum_2013,
  patch_kinetics_2017}.  Recent studies show that, unlike equilibrium
systems, these systems may not have an equation of state for the mechanical
pressure~\cite{solon_pressure_2015_1, solon_pressure_2015_2,
  junot_active_2017}.

The stochastic dynamics used to describe the motion of RTPs and other active
particles has been studied earlier in the context of systems with colored
noise, and some exact as well as approximate analytic results for steady-states and first-passage properties were
obtained~\cite{fox1986uniform,hanggi1995colored}.  The dynamics of active
particles is related to the equilibrium properties of semi-flexible polymers,
where many analytic results are known~\cite{dhar_triple_2002,supurna_2002}.  There have
been recent attempts to understand the time evolution of the probability
distributions of active particles in unbounded geometries
\cite{kurzthaler_intermediate_2016}.  In confined geometries, RTPs and ABPs
are known to accumulate near the boundaries of the domain
\cite{bechinger_active_2016}. The steady-state distribution of such active
particles in confining potentials are non-Boltzmannian
\cite{hanggi1995colored,Tailleur2008,argun_non-boltzmann_2016,das2017confined} and can
exhibit jammed states \cite{klein_spontaneous_2016, evans2016}.  {
  More recently there have been a number of studies on computing the steady-state distribution for both RTPs and ABPs in various confined geometries,
  but using approximate methods in most cases
  \cite{maggi_multidimensional_2015,wagner_steady-state_2017,
    angelani_confined_2017, elgeti2015run}.  However, so far, the approach to
  the steady-state has not been studied in detail. }

Given the rich behavior of RTPs, it is worthwhile to study them in the
simplest possible setting where we can derive explicit results for basic
dynamical observables.  In this spirit, we investigate the dynamics of
non-interacting RTPs with an additional Brownian diffusion term. We
investigate the motion on: (i) the infinite line, (ii) a one-dimensional
bounded domain with reflecting walls, and  (iii) the semi-infinite line and the bounded domain with absorbing boundaries.
The restriction to one dimension greatly simplifies the analysis, without sacrificing phenomenological
richness.

We  implement run and tumble motion by imposing a particle velocity $v$ that switches sign at a random Poisson rate.
Naively, one might anticipate that this velocity switching merely renormalizes the diffusion
coefficient.  Such an interplay between advection and diffusion underlies,
for example, the phenomenon of hydrodynamic
dispersion~\cite{Taylor186,Aris67,PhysRevLett.57.996,PhysRevA.37.2619}.
Here, a diffusing tracer is passively carried by a flow field, such as
Poiseuille flow in a pipe, and the combination of microscopic diffusion and
convection leads to a greatly enhanced spread of the tracer in the
longitudinal direction.  A similar phenomenon arises for RTPs in the
unbounded geometry in the long-time limit.  However, there are surprising
pre-asymptotic effects.  For a wide range of parameters, the probability
distribution evolves from unimodal, to multimodal, before finally
converging to a Gaussian in the long-time limit.  We also compute the 
steady-state of a RTP inside a finite domain, and examine at the approach to the
steady-state. The approach to the steady-state is studied by examining the
spectral structure of the relevant Fokker-Planck operator, and we find that
this problem is highly non-trivial. Finally we study first-passage properties
of the RTP inside a semi-infinite domain where we obtain exact analytic results for 
the first-passage distribution and exit time probabilities. We compare these results with the usual diffusive case and point out the qualitative differences.

This paper is organized as follows.  In section \ref{sec:model}, we define
the model and discuss the relevant boundary conditions for the probability
distributions in a finite interval.  In section \ref{sec:infiniteNonZeroD},
we calculate the propagators for RTPs with superimposed diffusion in an
unbounded domain and thereby derive the exact probability distribution.  We
study RTPs in a bounded domain and calculate their steady-state and
time-dependent distributions in section \ref{sec:bounded_domain}. Finally we
turn to first-passage properties of an RTP where we calculate its 
survival
probability in a semi-infinite one-dimensional domain with absorbing walls
(Sec.~\ref{sec:survivalProbability}) and the exit times in this domain
(Sec.~\ref{sec:exit_times}).  Throughout this work, we compare our exact
results with numerical simulations of the Langevin equations for the RTPs and
numerical solutions of the associated Fokker-Planck equation.

\section{RTP Model}
\label{sec:model}

We study a particle that moves on the one-dimensional line whose motion is
described by the following stochastic equation
\begin{eqnarray}
\frac{dx}{dt} = v \, \sigma(t) + \sqrt{2D} \, \eta(t)\,,
\label{eqn_Langevin}
\end{eqnarray}
where the random variable $\sigma(t)$ switches between $\pm 1$ at a Poisson
rate $\gamma$, and $\eta(t)$ is Gaussian white noise with
\begin{eqnarray}
\langle \eta(t) \rangle = 0, \qquad \langle \eta(t) \eta(t') \rangle =  \delta(t-t')~. 
\end{eqnarray}
Equation (\ref{eqn_Langevin}) can be reduced to a Markovian model if we specify the particle
state by both its position $x$ and its current velocity ($\pm 1$).  It is convenient to define $P_+(x,t)$ and $P_-(x,t)$ as the
probability density for the particle to be at position $x$ with velocities $+v$
and $-v$, respectively.  These state probabilities evolve according to the generalized form of telegrapher's equation
\begin{subequations}
\begin{align}
  \begin{split}
    \label{eqn:P}
 \partial_t P_{+} &= D \partial_x^2 P_{+} - v \partial_x P_{+} - \gamma \, P_{+} +  \gamma P_{-} \, , \\
\partial_t P_{-} &= D \partial_x^2 P_{-} + v \partial_x P_{-} + \gamma \, P_{+} -  \gamma P_{-} \, . 
\end{split}
\end{align}
This equation was perhaps derived first in the context of
electromagnetic theory~\cite{whittaker1910history} and was later derived in
several other contexts (see the review~\cite{weiss2002some} and the
references therein).  The probability $P(x,t)$ to find the particle at
position $x$ at time $t$ is the sum of the probabilities $P_{\pm}(x,t)$ of
finding the particle in the two states, i.e., $P = P_+ + P_-$.  We choose
$\gamma^{-1}$ as the unit of time and $v\gamma^{-1}$ as the unit of length to
recast \eqref{eqn:P} in the dimensionless form
\begin{align}
\label{eqn-P+-}
  \begin{split}
 \partial_t P_{+} &= \mathcal{D} \, \partial_x^2 P_{+} - \partial_x P_{+} - P_{+} + P_{-} \, ,
 \\
\partial_t P_{-} &= \mathcal{D} \, \partial_x^2 P_{-} + \partial_x P_{-} + P_{+} -P_{-} \,,
\end{split}
\end{align}
\end{subequations}
where the dimensionless diffusion constant $\mathcal{D}=D\gamma/v^2$ is 
the only parameter for the unbounded system.

For the finite interval $[-L,L]$, there is a second parameter: the
dimensionless interval length $\ell = L\gamma/v$.  When the particle is
restricted to a finite domain $x\in[-\ell, \ell]$, we impose the boundary
condition that when the particle hits the boundary, it stays stuck there until its internal state ($\pm$) changes, upon which it can move away from the boundary.  Hence there
is no particle current across these walls.  From Eqs.~\eqref{eqn-P+-}, we
identify the particle currents $J_{\pm}(x,t)$ at position $x$ and time $t$
as:
\begin{equation}
J_{\pm}(x,t)= -\mathcal{D} \partial_x P_{\pm} ~\pm ~  P_{\pm}. \label{J_pm}
\end{equation}
The following four boundary conditions are obtained by demanding that the
value of these currents is zero at $x=\pm \ell$, that is,
\begin{align}
  \begin{split}
    \label{BC12}
\left (\mathcal{D} \partial_x P_+  - P_+ \right)_{x=\pm \ell}&=0,  \\
\left ( \mathcal{D} \partial_x P_- + P_-\right)_{x=\pm \ell}&=0.
\end{split}
\end{align}

\section{The Occupation Probability $P(x,t)=P_+(x,t)+P_-(x,t)$:}
We now determine the RTP occupation probability $P(x,t)$, namely, the
probability that the particle is at position $x$ at time $t$ for: (a) the
infinite line and (b) the finite interval $[-\ell,\ell]$.  To compute $P(x,t)$, we
need to solve the coupled Fokker-Planck equations \eqref{eqn-P+-} for
$P_{\pm}(x,t)$ with the appropriate boundary conditions.

\subsection{Infinite domain: $\ell = \infty$}
\label{sec:infiniteNonZeroD}

It is useful to define the Fourier transforms
$ \tilde{P}_\pm(k)=\int_{-\infty}^\infty P_\pm(x,t) e^{i k
  x}~dx$. Fourier transforming Eqs.~\eqref{eqn-P+-} with respect to
$x$, we obtain (in matrix form):
\begin{align}
 & \frac{d}{dt} \begin{pmatrix} \tilde{P}_{+}(k,t) \\  \tilde{P}_{-} (k,t) \end{pmatrix} =
 \mathbb{A}_k
  \begin{pmatrix} \tilde{P}_{+} (k,t) \\  \tilde{P}_{-} (k,t) \end{pmatrix}\,,
  \label{Ppm_Eqn}
\end{align}
where
\begin{align}
  \mathbb{A}_k= \begin{pmatrix}
  -1-ik-\mathcal{D}k^2 & \gamma \\  \gamma & -1+ik-\mathcal{D}k^2 \end{pmatrix}\,.  \nonumber
\end{align}
Diagonalizing the matrix $\mathbb{A}_k$ for each $k$ and solving the
resulting linear equations gives
\begin{align}
  \begin{pmatrix} \tilde{P}_{+}(k,t) \\  \tilde{P}_{-} (k,t) \end{pmatrix} &=
 \mathbb{W}_k
  \begin{pmatrix} \tilde{P}_{+} (k,0) \\  \tilde{P}_{-} (k,0) \end{pmatrix}\,,
  \label{Ppm_Eqn-2}
\end{align}
where
\begin{align}
  \mathbb{W}_k &=  \begin{pmatrix}
\frac{e^{\alpha_+ t}}{\mathcal{N}_+^2} + \frac{e^{\alpha_-t}}{\mathcal{N}_+^2} \left( ik+\sqrt{1-k^2} \right)^2  & \frac{\gamma \left( e^{\alpha_+ t} - e^{\alpha_- t} \right)}{2 \sqrt{1 - k^2}}  \\ 
\frac{\gamma \left( e^{\alpha_+ t} - e^{\alpha_- t} \right)}{2 \sqrt{1 - k^2}}  & \frac{e^{\alpha_+ t}}{\mathcal{N}_-^2} + \frac{e^{\alpha_-t}}{ \mathcal{N}_-^2} \left( ik-\sqrt{1-k^2} \right)^2 \end{pmatrix}\,, \nonumber 
\end{align}
with $\mathcal{N}_\pm =\sqrt{2}~(1 -k^2 \pm ik \sqrt{1-k^2})^{1/2}$ and
$\alpha_{\pm}=-(1 +\mathcal{D}~k^2) \pm \sqrt{1-k^2}$.

Consider the natural initial condition in which the particle starts at $x=0$,
with equal probability to be  either in the $+$ or the $-$ state.  The
Fourier transform of the initial probability is then
$\tilde{P}(k,0)= (4 \pi)^{-1}$.  Using this in (\ref{Ppm_Eqn-2}) and
simplifying, we find
\begin{eqnarray}
  \label{eqn:Pk}
 \fl \qquad \tilde{P_{\pm}} (k,t) = \frac{e^{\alpha_+ t}}{2} 
  \left( \frac{1}{\mathcal{N}_\pm^2}+\frac{1}{2\sqrt{1-k^2}} \right) 
  + \frac{e^{\alpha_- t}}{2} 
  \left( \frac{\left( ik+\sqrt{1-k^2} \right)^2}{\mathcal{N}_\pm^2}-\frac{1}{2\sqrt{1-k^2}} \right). \label{Pp_infiniteline}
\end{eqnarray}
From \eqref{eqn:Pk}, the Fourier transform of the total probability
$\tilde{P}(k,t) = \tilde{P}_+(k,t)+\tilde{P}_-(k,t)$ is
\begin{equation}
  \label{totalP_infiniteline}
  \tilde{P} (k,t) = {e^{-\left(1 +\mathcal{D}k^2\right) t}}
  \left[  \textrm{cosh} \left( t\sqrt{1-k^2} \right)
  + \frac{1 }{\sqrt{1-k^2}}\,    \textrm{sinh} \left( t\sqrt{1-k^2}  \right)  \right]. 
\end{equation}

We can alternatively derive this result as follows: The displacement of an
RTP that starts at $x=0$, can be written formally by integrating the Langevin
equation (\ref{eqn_Langevin}) to give
$x(t) = \int_0^t \sigma(t) dt + \sqrt{2\mathcal{D}} \int_0^t \xi(t)
dt\equiv A(t) +B(t)$.  Since the random processes $A(t)$ and $B(t)$ are
independent of each other, 
$\tilde{P}(k,t) = \left\langle e^{ikA(t)} \right\rangle \left\langle
  e^{ikB(t)} \right\rangle$.  It is easy to see that the expression in
(\ref{totalP_infiniteline}) is actually in this product form once one
identifies $\left\langle e^{ikB(t)} \right\rangle = e^{-\mathcal{D} k^2 t}$
for the Brownian motion $B(t)$. The process $A(t)$ is the motion of an RTP
with $\mathcal{D}=0$, whose dynamics is described by the telegrapher's
equation, for which $\left\langle e^{ikA(t)} \right\rangle$ can be computed
explicitly (see e.g., \cite{dhar_triple_2002,supurna_2002}).  This immediately leads to
the expression in (\ref{totalP_infiniteline}).

Using the product structure of $\tilde{P}(k,t)$, we invert the Fourier
transform in (\ref{totalP_infiniteline}) to derive the probability $P(x,t)$
in the convolution form $P(x,t)= \int_{-\infty}^\infty g(x-y,t)h(y,t)$ where
$g(x,t)= \exp(-x^2/4\mathcal{D}t)/\sqrt{4 \pi \mathcal{D} t}$ is the inverse
Fourier transform of
$\left\langle e^{ikB(t)} \right\rangle = e^{-\mathcal{D} k^2 t}$ and $h(x,t)$
is the inverse Fourier transform of $\left\langle e^{ikA(t)}
\right\rangle$. Using the explicit expression of $h(x,t)$ from
\cite{dhar_triple_2002,supurna_2002}, 
we obtain
\begin{align}
P(x,t)&=\frac{\textrm{cosh}(x/2\mathcal{D})}{\sqrt{4 \pi \mathcal{D} t}}e^{-t-(x^2+t^2)/4\mathcal{D}t} \nonumber \\ 
& + \frac{e^{-t}}{2} \int_{-\infty}^\infty \!\!dy \, \frac{e^{-(x-y)^2/4\mathcal{D}t}}{\sqrt{4 \pi \mathcal{D} t}}\left [I_0\left (\sqrt{t^2-y^2} \right) +\frac{t}{\sqrt{t^2-y^2}}I_1\left(\sqrt{t^2-y^2}\right) \right ]\Theta(t-|y|),
\label{p-inf}
\end{align}
where $I_{n}$ is the $n^{\rm th}$-order modified Bessel function of the
first kind and $\Theta$ is the Heaviside step function. Note that in the
limit $x,t \to \infty$, $P(x,t)$ reduces to a simple Gaussian with diffusion
constant $(\mathcal{D}+1/2)$. This can be easily seen from
(\ref{totalP_infiniteline}) where
$\tilde{P}(k,t) \to \exp [- (\mathcal{D}+1/2)k^2 t ]$ as $t \to \infty$ and
$k\to 0$.
  
It is instructive to examine the spatial moments of the probability
distribution. All odd moments are zero by symmetry. Formally, the even moments of the distribution are given by
\begin{equation*}
\langle x^{2n}(t)\rangle = (-1)^n\frac{\partial^{2n} \tilde{P}(k,t)}{\partial k^{2n}}\Big|_{k=0}\,.
\end{equation*}
For the second moment, we  find
\begin{equation}
\label{xt}
\langle x^{2}(t)\rangle = (2\mathcal{D} + 1) \, t \; - \frac{(1-e^{-2 t})}{2}\,
\to
(2D + v^2/\gamma) \, t \; - \frac{v^2}{2\gamma^2} (1-e^{-2 \gamma t})  .
\end{equation}
where in the last simplification we have put in all the dimensional parameters. The above result has two non-trivial limiting cases.  For $\gamma \neq 0$ and
$t\to\infty$, \eqref{xt} reduces to
\begin{equation}
\label{xta}
\langle x^{2}(t)\rangle\simeq (2\mathcal{D}+1)t \to  (2D+v^2/\gamma)t\,.
\end{equation}
In the
$t\to\infty$ limit, the finite switching rate $\gamma$ leads to an enhancement of the
microscopic diffusion coefficient in a manner that is reminiscent of
hydrodynamic
dispersion~\cite{Taylor186,Aris67,PhysRevLett.57.996,PhysRevA.37.2619}.  On
the other hand, in the limit $\gamma\to 0$, we find
\begin{equation}
\langle x^{2}(t)\rangle\to 2\mathcal{D}t + v^2 t^2\,.
\end{equation}
Thus the mean-square displacement crosses over from growing linearly with $t$
to quadratically with $t$ as $\gamma\to 0$.

We can also compute higher-order derivatives of $\tilde{P}(k,t)$ from which higher moments of the displacement can be deduced.  The fourth moment is
\begin{align}
\langle x^4(t)\rangle &= 3t^2\left(2D + \frac{v^2}{\gamma}\right)^2 -
\frac{3v^2 \, t}{\gamma^3}\left[2D\gamma \left(1 - e^{-2\gamma
  t}\right) - v^2\left(2 + e^{-2\gamma t}\right)\right]
+\frac{9 v^2}{2\gamma^4}\left(1 - e^{-2\gamma t}\right).
\end{align}
The important feature of this last result is that as $t\to\infty$,
$\langle x^4\rangle/ 3 \langle x^2\rangle^2\to 1$, which is just the relation
between the fourth and second moments for a Gaussian distribution.  The
behavior of the higher moments also conforms to those of the Gaussian
distribution as $t\to\infty$.

\begin{figure}
\centering
\includegraphics[width=\textwidth]{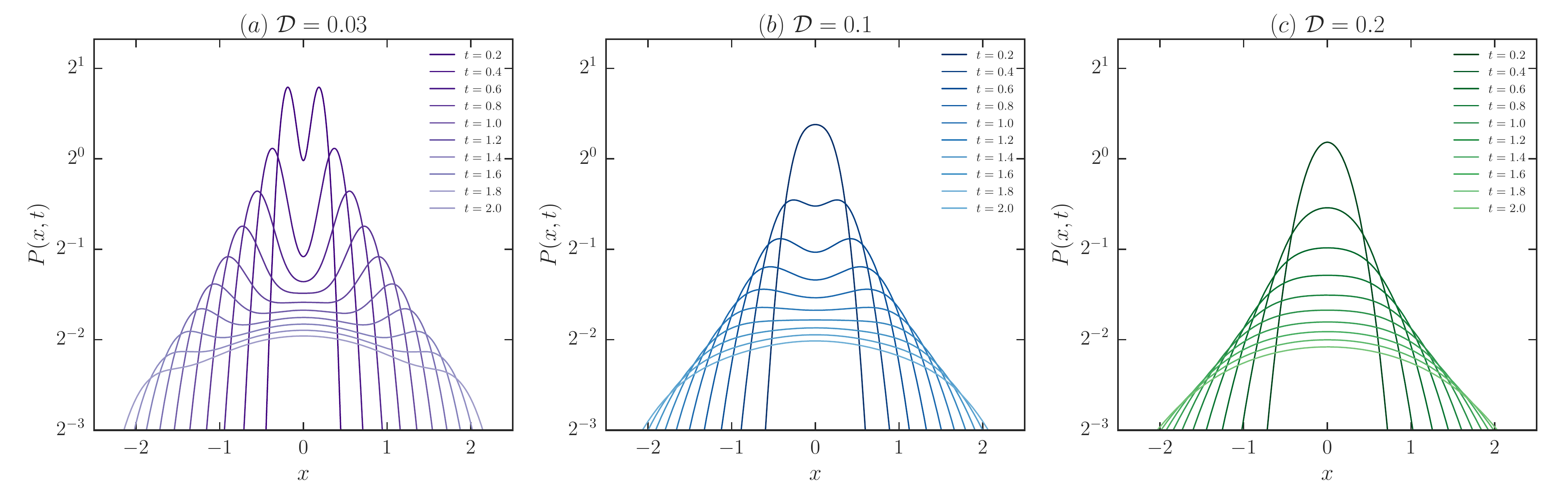}
\caption{ Plot of the probability density $P(x,t)$ in (\ref{p-inf}) for three different values of the diffusion constant $\mathcal{D}$.}
\label{fig:time-evolution-unbounded}
\end{figure}

In Fig.~\ref{fig:time-evolution-unbounded}, we plot the temporal evolution of
this occupation probability $P(x,t)$ for different values of the
dimensionless diffusion coefficient $\mathcal{D}$. For $\mathcal{D}$ greater
than a critical value $\mathcal{D}_c$, the probability distribution is
unimodal for all times. However, for
$\mathcal{D} < \mathcal{D}_c~(\approx 0.175)$, the occupation probability
evolves from a unimodal distribution at short times, to a {multimodal}
distribution at intermediate times and finally back to a unimodal
distribution at long times. This non-trivial behavior of $P(x,t)$ for small
$\mathcal{D}$ arises from the competition between the stochastic flipping of
particle states (at rate $\gamma$) and translational diffusion. For a small
$\mathcal{D}$, since $P_{\pm}(x,0)=\delta(x)/2$, the RTPs in the $+$ ($-$)
states move to the right (left) in an almost ballistic manner. This splits
the initial unimodal distribution into a bimodal distribution with two
symmetric peaks (see Fig.~\ref{fig:time-evolution-unbounded}). While these
two peaks are moving ballistically in opposite directions, they are also
broadening because of the true diffusion term. As a result, at 
intermediate times when the tails of these two separated peaks meet at the
centre, there again starts accumulation of particles (see
Fig.~\ref{fig:time-evolution-unbounded}(a)). This may lead to a central peak
before the two ballistically moving side peaks disappear, which depends on the
relative strengths of $\gamma$ and $\mathcal{D}$. Once developed, the central
peak starts continuously broadening and on time scales much longer than the
stochastic flipping rate $\gamma^{-1}$, the RTPs remix, leading to an
effective diffusion constant as discussed above. As a result, the multimodal
distribution at intermediate times merges into a unimodal distribution, which
as $t\to \infty$, converges to a Gaussian. On the other hand, for large
$\mathcal{D}$, the split peaks of the two RTP states overlap to such an
extent that the full distribution always remains unimodal.  This behavior
suggests that there exists a critical $\mathcal{D}$ where the effects of
translational diffusion and stochastic flipping balance each other.

\begin{figure}
\centering
\includegraphics[width=0.9\textwidth]{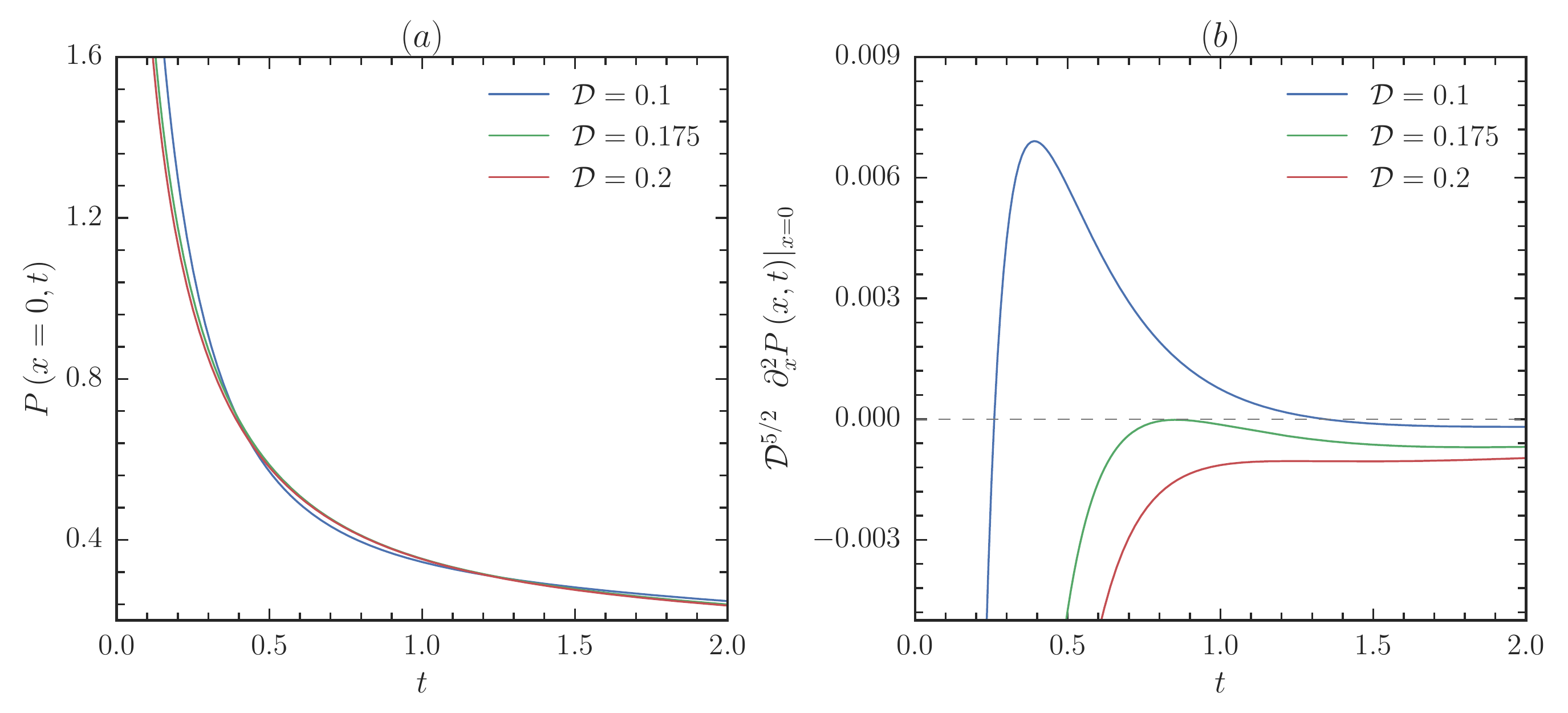}
\caption{(a) The time-evolution of occupation probability $P(x,t)$ at $x=0$
  for three-different values of the dimensionless diffusion coefficient
  $\mathcal{D}$. The occupation probability $P(0,t)$ is higher for smaller
  values of $\mathcal{D}$ at early-times, is lower at intermediate times and
  is again higher at long times. (b) The nature of the extremum of $P(x,t)$
  at $x=0$ changes from being a maximum at earlier times to a minimum at
  intermediate times, and finally to a maximum at long times. This
  non-trivial behavior of the central extrema exists only for
  $\mathcal{D} < \mathcal{D}_{c} \approx 0.175$. For $\mathcal{D} \geq \mathcal{D}_c$, $P(x=0,t)$ is always a maxima
  for all times.}
\label{fig:unbounded-central-time-evolution}
\end{figure}

To understand this transition, we plot $P(x=0,t)$ for various $\mathcal{D}$
values in Fig.~\ref{fig:unbounded-central-time-evolution}(a), and notice that
the occupation probability at $x=0$ is higher for smaller $\mathcal{D}$ at
short times as compared to that for larger $\mathcal{D}$. This then crosses
over to a lower value at intermediate times and finally becomes larger at
long times. Furthermore, to investigate the nature of the occupation
probability at $x=0$, we plot the second derivative
$\partial_x^2 P(x,t) \vert_{x=0}$. We find that for small $\mathcal{D}$,
$P(x=0,t)$ has a maximum at short times, crosses over to a minimum at
intermediate times, and finally crosses over to a maximum again at long
times.  However, for the critical value of $\mathcal{D}_{c}\approx 0.175$, these two
crossover times merge, resulting in a unimodal distribution. For
$\mathcal{D} \geq 0.175$, we find that $\partial_x^2 P(x,t) \vert_{x=0}$ is always
negative and hence $P(x,t)$ is always unimodal.

In summary, we find that, in contrast to the Gaussian form for a Brownian particle, 
the probability
distribution for an RTP can be multimodal depending on the value of the
dimensionless diffusion coefficient $\mathcal{D}$.  This diversity in the
probability distribution also occurs in other systems in which the motion of
a diffusing particle is influenced by an interplay with a convection field
that changes sign~\cite{PhysRevA.45.7207}.

\subsection{Bounded interval}
\label{sec:bounded_domain}

We now treat an RTP in the interval $x \in [-\ell,\ell]$.  In this case, the
probability distribution will reach a steady-state in the long time limit.
For $v=0$, the particle performs pure Brownian motion and reaches a spatially
uniform steady-state at long times.  On the other hand if $\mathcal{D}=0$,
the particle is subjected to only the dichotomous noise $\sigma(t)$.  Here,
the particle reaches a different steady-state in which, for any finite
flipping rate, there is an accumulation of particles at the
boundaries. However, for very large flipping rates, one regains a spatially
uniform distribution with an diffusion effective coefficient $v^2/\gamma$.
In the case where both $v$ and $\mathcal{D}$ are nonzero, we anticipate a
steady-state which is intermediate to these two extreme cases.  We first
solve for the probability distribution in the steady-state and then we turn
to the more complicated time-dependent solution.

In the steady-state, the dimensionless Fokker-Planck equations \eqref{eqn-P+-} reduce to
\begin{align}
  \label{eqn:Pss}
  \begin{split}
&\mathcal{D} \, \partial_x^2 P_{+} - \partial_x P_{+} - P_{+} + P_{-} =0 \,, \\
&\mathcal{D} \, \partial_x^2 P_{-} + \partial_x P_{-} + P_{+} -P_{-} =0 \,.
\end{split}
\end{align}
which we have to solve subject to the boundary conditions (\ref{BC12}).  The
details of this calculation are given in \ref{app:interval}. The final result for the probability distribution is:
\begin{align}
P(x)={\left[\frac{\tanh\left(\frac{\sqrt{2\mathcal{D}+1} }{\mathcal{D}} \ell\right)}{\sqrt{2\mathcal{D}+1}} +2\ell\right]}^{-1}\left[\frac{\cosh\left(\frac{\sqrt{2\mathcal{D}+1}}{\mathcal{D}} x\right)}{2 \mathcal{D} \cosh\left(\frac{\sqrt{2\mathcal{D}+1}}{\mathcal{D}} \ell \right)}  +1\right]\,.
\label{expression_P}
\end{align}

\begin{figure}[ht]
\centering
 \includegraphics[width=\textwidth]{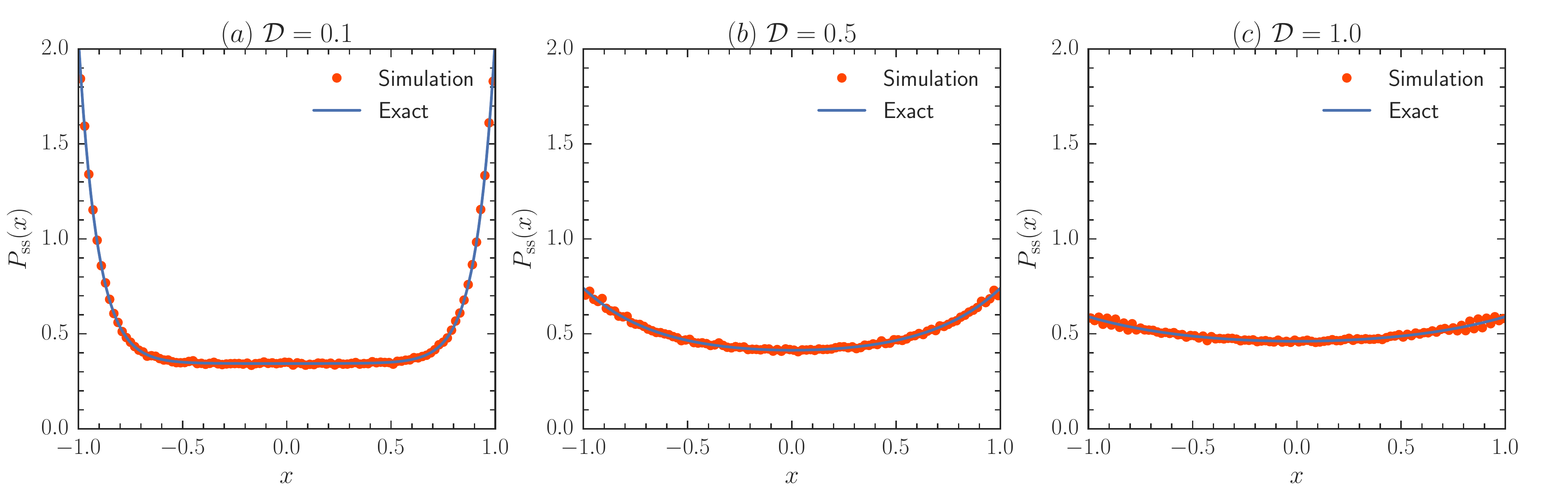}
 \caption{Comparison of the steady-state probability density equation
   \eqref{expression_P} with explicit Langevin simulations for various
   $\mathcal{D}$. The histogram of the numerical simulation was constructed,
   at $t=5$, using $10^6$ different realizations of the stochastic process.}
\label{fig:bounded-steady-state}
\end{figure}

In Fig.~\ref{fig:bounded-steady-state}, we compare \eqref{expression_P} for
the steady-state probability distribution $P(x)$ with results of simulation
of the Langevin equation~\eqref{eqn_Langevin}, and find nice agreement. We
observe that probabilities are higher near the boundaries than at the center
of the interval, in contrast to the uniform density which one would observe if
there was no activity \emph{i.e.} $v=0$. Such accumulation of active
particles near the boundaries of a confined domain is quite generic and has
been observed in experimental systems such as motile rods~\cite{bricard_2013}
and bacterial suspensions~\cite{di_leonardo_bacterial_2010}.

In the limit $\mathcal{D} \to 0$, the peaks near the
boundaries become progressively sharper, and eventually become
delta-function peaks. The full distribution is given by
\begin{eqnarray}
  P(x)|_{\mathcal{D} \to 0} =\frac{2 + \delta(x-\ell)+\delta(x+\ell) }{2(1 + 2\ell)}~.
\end{eqnarray}
We observe that the probability is uniform everywhere except for the delta
function peaks at the boundaries.  This $\mathcal{D} \to 0$ case has recently
been considered in a similar context~\cite{angelani_confined_2017} and our
method reproduces their results.

\begin{figure}[ht]
\centering
 \includegraphics[width=0.7\textwidth]{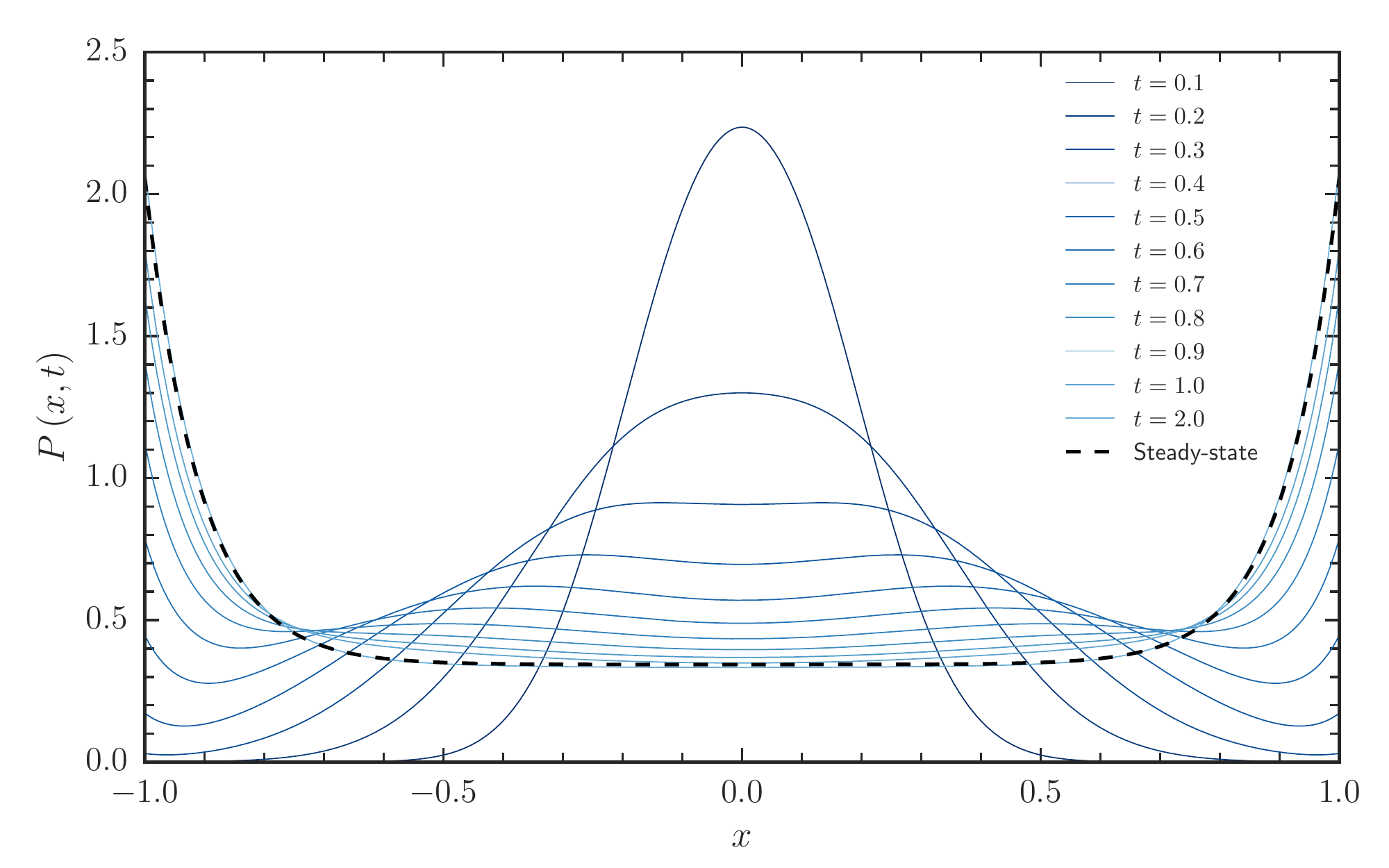}
 \caption{Time evolution of $P(x,t)$ in a interval obtained from solving the
   Fokker-Planck equations with appropriate boundary conditions. The
   diffusion constant was set to $\mathcal{D}=0.1$ and the data corresponds
   to an initial condition $P_+(x,0)=P_-(x,0)=\delta(x)/2$. It can be seen
   that at late times, the distribution converges to the exact steady-state
   distribution \eqref{expression_P}.}
\label{fig:bounded-time-evolution}
\end{figure}

Let us now turn to the full time-dependent solution.  We are interested in
how the distribution $P(x,t)$ approaches the steady-state in the
$t \to \infty$ limit.  To this end, we have to solve the coupled
time-dependent Fokker-Planck equations \eqref{eqn-P+-} within the interval
$x \in [-\ell,\ell]$, subject to the boundary conditions (\ref{BC12}).  For
the time-dependent solution, we expand $P(x,t)$ in terms of the complete set
of basis functions as
\begin{equation}
\begin{pmatrix} P_+(x,t) \\ P_-(x,t)\end{pmatrix} = \sum_{n} a_n e^{\lambda_n t}~
\begin{pmatrix} \phi^+_n(x) \\ \phi^-_n(x) \end{pmatrix}\,, \label{seriessol}
\end{equation}
where $\lambda_n$ are the eigenvalues and the eigenfunctions
$[\phi^+_n(x),\phi^-_n(x)]$ satisfy
\begin{align}
  \label{eqn:lambda}
  \begin{split}
&    \mathcal{D}~\partial_{xx} {\phi}^+_{n} -  \partial_x {\phi}^+_{n}  -
    {\phi}^+_{n} +   {\phi}^-_{n} = \lambda_n {\phi}^+_{n}\,, \\
& \mathcal{D}~\partial_{xx} \phi^{-}_n +  \partial_x {\phi}^{-}_n  + {\phi}^{+}_n -  {\phi}^{-}_n
= \lambda_n {\phi}^-_{n}~\,,
  \end{split}
\end{align}
subject to the boundary conditions (\ref{BC12}).  The coefficients $a_n$ are
given in terms of the left eigenvectors $\langle \chi_n|=[\chi_n^+(x),\chi_n^-(x)]$ as
\begin{equation}
a_n=\langle \chi_n|P(x,t=0)\rangle=\int_{-\ell}^\ell dx~ \left[~\chi_n^+(x) P_+(x,t=0)+  \chi_n^-(x) P_-(x,t=0)~\right]~.
\end{equation}
The left eigenvectors can be obtained as solutions of
the adjoint Fokker-Planck operator. It can be shown that this has the same form as that in Eqs.~\eqref{eqn:lambda}, with the sign of the
$\partial_x$ term changed, and with Neumann boundary conditions for both
$\chi_n^+(x)$ and $\chi_n^-(x)$.

The details of calculating the eigenstates $\phi_n^\pm(x)$ are given in
\ref{app:interval-td}.  Here we compare the time-evolution of the probability
density obtained from a numerical solution of the Fokker-Planck equations
\eqref{eqn-P+-} using the boundary conditions \eqref{BC12}, with the spectral
expansion given by (\ref{seriessol}).  The ground state eigenvalue
$\lambda_0=0$, and the corresponding eigenstate (the steady-state) is known
exactly and given by Eq.~(\ref{expression_P}).  In
Fig.~\ref{fig:bounded-time-evolution}, we show the time-evolution of $P(x,t)$
obtained from a numerical solution of equations \eqref{eqn-P+-}. The unimodal
to bimodal crossover discussed in the unbounded system can be seen in the
figure. We also observe that our numerical solution converges to the exact
steady-state.
\begin{figure}[ht]
\centering
\includegraphics[width=\textwidth]{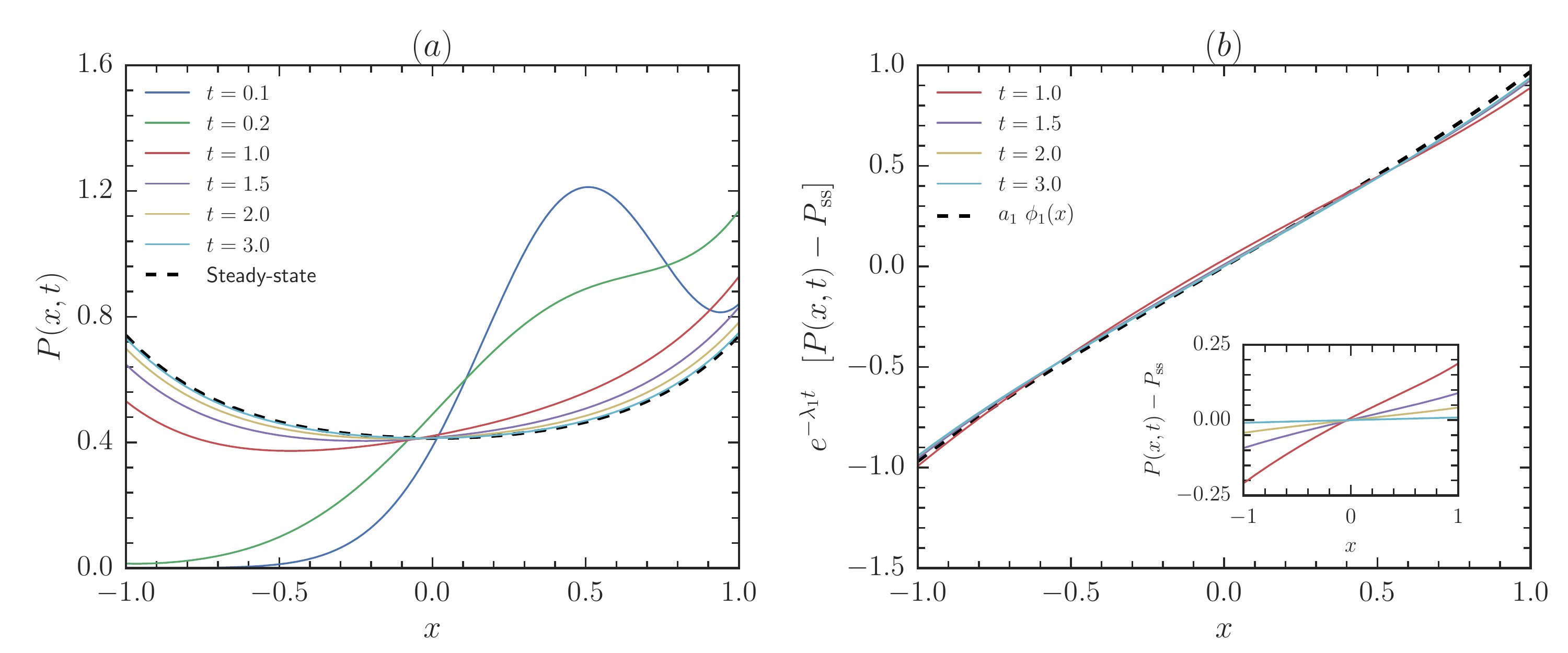}
\caption{(a) Time evolution of $P(x,t)$ in an interval obtained from solving the
  Fokker-Planck equations with appropriate boundary conditions. The diffusion
  constant was set to $\mathcal{D}=0.5$ and the data corresponds to an
  initial condition $P_+(x,0)=P_-(x,0)=\delta(x-1/2)/2$. It can be seen that
  at long times, the distribution converges to the exact steady-state
  distribution. (b) Comparison of $ (e^{-\lambda_1 t} [ P(x,t)-P_{SS}(x)]$
  with the eigenstate $\phi(x)$ corresponding to first excited state. The
  initial state here was chosen as $P_\pm(x,t=0)=\delta{(x-1/2)}/2$, same as
  in (a), which has $a_1 \approx 0.7789$. The inset shows the unscaled data.}
\label{fig:bounded-time-evolution-comparison}
\end{figure}
All the other eigenvalues $\lambda_n,~ n>0$ have to be found numerically from
the zeros of the determinant of the matrix ${M}$ in (\ref{c-eq}).  The
long-time relaxation of the system to the steady-state would be determined by
the eigenvalue with the largest non-zero real part, and the corresponding
eigenfunction. As an illustrative example, we choose $\mathcal{D}=0.5$ and
$\ell=1$, in which case the dominant eigenvalue is given by
$\lambda_1 \approx -1.55684$, while the corresponding eigenfunction is given
by (\ref{phisoln}), with
\begin{align}
&(\beta^{(1)}_{++},\beta^{(1)}_{-+},\beta^{(1)}_{+-},\beta^{(1)}_{--}) = (1.66409,-1.66409,-0.998284 i, 0.998284 i)~, \nonumber \\
& (\alpha^{(1)}_{++},\alpha^{(1)}_{-+},\alpha^{(1)}_{+-},\alpha^{(1)}_{--})~ \nonumber \\
&=(-0.277351, -3.60554, -0.0585543 - 0.998284 i, -0.0585543 + 0.998284 i)~, \nonumber \\
&(C^{(1)}_{++},C^{(1)}_{-+},C^{(1)}_{+-},C^{(1)}_{--}) \nonumber\\
&= (0.169039, 0.0468831,-0.196156 + 0.184987 i,-0.196156 - 0.184987 i)~. \nonumber
\end{align}
Thus we know the functions $\phi_1^+(x)$, $\phi_1^-(x)$ explicitly, and in terms of these,  we expect at long times
\begin{equation}
P(x,t)=P_+(x,t)+P_-(x,t)=P_{SS}+ a_1 e^{\lambda_1 t} [\phi_1^+(x)+\phi_1^-(x)]+\ldots~.\label{spectralR}
\end{equation}
The parameter $a_1$ depends on initial conditions and can be obtained from
the corresponding left eigenvector $\langle \chi_1 |$ and we find
$a_1=\langle \chi|P(t=0)\rangle=0.7789...$.  In
Fig.~(\ref{fig:bounded-time-evolution-comparison}a), we plot the evolution of
$P(x,t)$ obtained from a direct numerical solution of Eqs.~\eqref{eqn-P+-}
and \eqref{BC12}, starting from an initial condition $\delta(x-1/2)$. The
plot in Fig.~(\ref{fig:bounded-time-evolution-comparison}b) shows
$P(x,t)-P_{SS}$ and compares this with the prediction from the first term in
the spectral representation Eq.~\eqref{spectralR}. It is seen that agreement
is very good. In general, we find that eigenvalues can have imaginary parts
(in which case they come in complex conjugate pairs) and so one can see
oscillatory relaxation.

\section{First-Passage Properties}

In biological systems, we are often interested in the time required for a molecule
diffusing in the interior of the cell to get adsorbed at the cell boundaries,
as well as in the time required by a diffusing protein to find the correct
binding sites.  Similarly, in chemical reactions, an important quantity
is the time spent by a reactive agent before it reaches catalytic boundaries.
Hence it is important to compute quantities such as first-passage
distributions, survival probabilities, and exit time distributions for the 
RTP.
First-passage and survival probabilities of stochastic processes have
been widely studied in the past (for reviews see 
\cite{R01,review_persistence}). In the context of RTP in one dimension, 
with the position evolving via Eq.
(\ref{eqn_Langevin}) but without the diffusion term, i.e., for $D=0$, 
first-passage properties have been studied 
before~\cite{weiss2002some}. More recently, the mean first-passage time
between two points in space was computed for RTP (again for $D=0$)  
analytically~\cite{ADP2014}. For a RTP in one dimension, the mean 
first-passage time was recently measured 
numerically~\cite{scacchi2017mean}.
In this section we study the first-passage probability analytically
for an RTP on a semi-infinite line and exit problem from a finite interval, in presence
of telegraphic as well as the diffusive noise, i.e, when both terms
in Eq. (\ref{eqn_Langevin}) are present.

\subsection{First-passage on the semi-infinite line}
\label{sec:survivalProbability}

We are interested in the probability for an RTP, which starts from a point $x$
on the semi-infinite line with velocity $\pm 1$, to arrive at the origin for the first time at time
$t$.  This quantity is directly related to the survival probability of the
RTP in the same geometry in the presence of an absorbing
boundary at $x=0$.  Let $S_+(x,t)$ [$S_-(x,t)$] denote the probability that
the RTP, starting initially at $x\ge 0$ with a positive [negative] 
velocity, survives being absorbed at the origin $x=0$ until time $t$, 
i.e.,
it does not cross the origin up to time $t$. Given the Langevin equation
\eqref{eqn_Langevin}, it is convenient to write the backward Fokker-Planck 
equations for the evolution of $S_{\pm}(x,t)$, where the initial position
$x$ is treated as a variable~\cite{bfp_review}. Consider the evolution 
over the time window 
$[0,t+dt]$ and break it into two sub-intervals $[0,dt]$ and $[dt, t+dt]$. 
It follows from Eq.~\eqref{eqn_Langevin} 
that in a small time $dt$ following $t=0$ (i.e., during the first 
interval $[0,dt]$), the 
position of the particle evolves to a new 
position $x'= x+ v\sigma(0)\, dt + \sqrt{2 {D}}\eta(0)\, dt$, where 
$\sigma(0)$ and $\eta(0)$
are the initial noises. For the subsequent evolution in the time interval 
$[dt, t+dt]$, the
new starting position is then $x'$.  Thus the survival probability
satisfies the evolution equations
\begin{align}
  \label{nbfp}
  \begin{split}
S_{+}(x, t+dt) & = (1-\gamma dt) \langle S_+(x+ v\, dt+ \sqrt{2D}\,\eta(0)\, 
dt, t)\rangle + \gamma dt~S_-(x,t)\,,    \\
S_{-}(x,t+dt) & =  (1-\gamma dt) \langle S_-(x- v\, dt+ \sqrt{2D}\,\eta(0)\, 
dt, t)\rangle + \gamma dt~S_+(x,t)\,, 
\end{split}
\end{align}
where the $\langle\, \rangle$ denotes the average over $\eta(0)$.  Expanding
in Taylor series for small $dt$, using the properties of $\eta(0)$ and taking
$dt\to 0$ limit, one directly arrives at a pair of backward equations which
read, in dimensionless units,
\begin{align}
\label{eqnQ}
\begin{split}
\partial_t S_{+}(x,t)= - S_{+}+ S_{-} +  \partial_x S_{+} + \mathcal{D}\, 
\partial_x^2 S_{+}\,,\\
\partial_t S_{-}(x,t)= \phantom{-} S_{+} -   S_{-} -  \partial_x S_{-} +
\mathcal{D}\, \partial_x^2 S_{-}\,.
\end{split}
\end{align}

These equations are valid for $x\ge 0$, with the initial conditions
$S_{\pm}(x,0)= 1$ for all $x>0$. In addition, we need to specify the boundary
conditions. As the starting point $x\to \infty$, it is clear that
$S_{\pm}(x\to \infty, t)=1$, since the particle will surely not cross the
origin in a finite time.  In contrast, the boundary condition at $x=0$ is
subtle: it depends on whether ${\cal D}=0$ or ${\cal D}>0$. Consider first
the case ${\cal D}=0$, i.e., in absence of normal diffusion. In this case, if
the particle starts at $x=0$ with a negative velocity, it will surely cross
the origin in a finite time.  Hence
\begin{subequations}
\begin{equation}
S_{-}(x=0,t)=0 \quad\quad {\rm when} \quad {\cal D}=0 \, .
\label{cald0}
\end{equation}
However, note that if the particle starts with a positive velocity, it can
survive up to finite $t$, hence the boundary condition $S_+(0,t)$ is
unspecified. We will see below that just one boundary condition in
Eq.~\eqref{cald0} is sufficient to make the solution of Eqs.~\eqref{eqnQ}
unique. Under normal diffusion, it is well known that if a particle
crosses the origin at some time, it recrosses it immediately infinitely often~\cite{review_persistence}. Hence, if the particle starts at the origin,
no matter whether the initial velocity is positive or negative, it will
surely cross zero within a short time $dt$, provided ${\cal D}>0$.
This follows from the fact that in the $dt \to 0$ limit, the Brownian noise dominates over the drift term irrespective of its sign. Hence, in
this case, we have the two boundary conditions
\begin{equation}
S_{\pm}(0, t) =0 \quad\quad {\rm when} \quad {\cal D}>0 \, . 
\label{caldnonzero}
\end{equation}
\end{subequations}
We will see later that indeed for ${\cal D}>0$, we will need both boundary
conditions in Eq.~\eqref{caldnonzero} to fix the solutions of
Eq.~\eqref{eqnQ} uniquely.

It is convenient to first take a Laplace transform, with respect to time 
$t$, of the pair of equations \eqref{eqnQ}.  Using the initial conditions
$S_{\pm}(x, 0)=1$, it is easy to see that the Laplace transforms
satisfy 
\begin{align}
  \label{lbfp}
  \begin{split}
    -1 + s\, \tilde{S}_{+}(x,s) &= {\cal D}\, \partial_x^2 \tilde{S}_{+} +
    \partial_x \tilde{S}_{+} - \tilde{S}_{+}+ \tilde{S}_{-} \,, \\
    -1 + s\, \tilde{S}_{-}(x,s) &= {\cal D}\, \partial_x^2 \tilde{S}_{-} -
    \partial_x \tilde{S}_{-} - \tilde{S}_{-}+ \tilde{S}_{+} \,,
\end{split}
\end{align}
where $\tilde{S}_{\pm}(x,s)= \int_0^{\infty} dt\, e^{-st}\, S_{\pm}(x,t)$
is the Laplace transform.

These equations can be made homogeneous by the 
shift: $\tilde{S}_{\pm}(x,s)= 1/s + U_{\pm}(x,s)$, where $U_{\pm}$ satisfy 
\begin{align}
  \label{upm1}
  \begin{split}
    \left[{\cal D}\partial_x^2 + \partial_x - (1+s)\right] U_{+}(x,s) & = -
    U_{-}(x,s)\,,\\
\left[{\cal D}\partial_x^2 - \partial_x - (1+s)\right] U_{-}(x,s) & = 
- U_{+}(x,s)\,.
\end{split}
\end{align}
Furthermore, by differentiating twice, one can write closed equations for 
$U_{+}$ and $U_{-}$
\begin{align}
  \label{upm2}
  \begin{split}
    \left[{\cal D}\partial_x^2 - \partial_x - (1+s)\right]\left[{\cal 
D}\partial_x^2 + \partial_x - (1+s)\right]U_+(x,s)= U_+(x,s) \,,\\
\left[{\cal D}\partial_x^2 + \partial_x - (1+s)\right]\left[{\cal
D}\partial_x^2 - \partial_x - (1+s)\right]U_-(x,s)= U_-(x,s)\,.
\end{split}
\end{align}
Below, we first solve the simpler case ${\cal D}=0$, followed by the
more complex ${\cal D}>0$ case.

\subsubsection{The case ${\cal D}=0$.}

This particular case has been considered earlier with space-dependent
transition rates, where only the mean first-passage time was
computed~\cite{masoliver1992solutions}.  Here we are interested in the full
first-passage time distribution, which we can obtain using the above backward
Fokker-Plank equation approach.  For ${\cal D}=0$, Eqs.~\eqref{upm2} are
ordinary second-order differential equations with constant
coefficients. Hence, we can try solutions of the form:
$U_{\pm}(x,s)\sim e^{-\lambda x}$.  Substituting this in either of
Eqs.~\eqref{upm2}, we find that $\lambda$ satisfies the quadratic equation,
$(\lambda+1+s)(\lambda-1-s)=1$, which gives two roots:
$\lambda(s)= \pm \sqrt{s^2+2s}$. Obviously, the negative root is not
admissible, since the solution must remain finite as $x\to
\infty$. Retaining only the positive root, the general solutions of
Eqs.~\eqref{upm2} can be written as
\begin{equation}
U_{+}(x,s) = B\, e^{-\lambda(s)\, x};\quad 
U_{-}(x,s) = A\, e^{-\lambda(s)\, x}\quad\, {\rm where}\quad 
\lambda(s)=\sqrt{s^2+2s}\,.
\label{AB}
\end{equation}
The two unknown constants $B$ and $A$ are however related, as they must also
satisfy the pair of first-order equations \eqref{upm1} (upon setting
${\cal D}=0$). This gives $B= A/[1+s+\lambda(s)]$. Hence, finally, using
$\tilde{S}_{\pm}(x,s)=1/s+ U_{\pm}(x,s)$, we get
\begin{align}
  \label{spm1}
  \begin{split}
\tilde{S}_{+}(x,s) &=  \frac{1}{s}+ \frac{A}{1+s+\lambda(s)}\, 
e^{-\lambda(s)\, x}\,,\\
\tilde{S}_{-}(x,s) &=  \frac{1}{s}+ A\,
e^{-\lambda(s)\, x}\,,
\end{split}
\end{align}
where $\lambda(s)=\sqrt{s^2+2s}$.
It remains to fix the only unknown constant $A$. This is done by using
the boundary condition $\tilde{S}_{-}(0,s)= 0$ which fixes $A=-1/s$.
Hence, we obtain the final solutions
\begin{align}
  \label{spm2}
  \begin{split}
\tilde{S}_{+}(x,s) &=  \frac{1}{s}\left[1- \frac{1}{1+s+\lambda(s)}\, 
e^{-\lambda(s)\, x}\right] \,, \\
\tilde{S}_{-}(x,s) &=  \frac{1}{s}\left[1- 
e^{-\lambda(s)\, x}\right]\,,
\end{split}
\end{align}
with $\lambda(s)=\sqrt{s^2+2s}$.

The first-passage time distribution is simply related to the survival
probability via
\begin{eqnarray}
f_+(x,t)=-\partial_tS_+(x,t)\,,\qquad   f_-(x,t)=-\partial_tS_-(x,t)\,,
\end{eqnarray}
or in Laplace variables
\begin{align}
  \label{fpm1}
  \begin{split}
\tilde{f}_+(x,s)&= 1-s \tilde{S}_+(x,s)=  \frac{1}{(s+1+\lambda(s))} 
\,\,e^{-\lambda(s) x}\,,  \\
\tilde{f}_-(x,s)&= 1-s \tilde{S}_-(x,s)=e^{-\lambda(s) x}~\,,
\end{split}
\end{align}
where we recall $\lambda(s)=\sqrt{s^2+2s}$. 

It turns out that the Laplace transforms in Eqs.~\eqref{fpm1} can be
exactly inverted.  Before doing so, it is useful to extract the long-time
asymptotics directly from the Laplace transforms in Eqs.~\eqref{fpm1}, by
considering the $s\to 0$ limit. A scaling limit then naturally emerges where
$s\to 0$, $x\to \infty$ but with the product $x\sqrt{s}$ fixed. This
corresponds, in the time domain, to the scaling limit $t\to \infty$,
$x\to \infty$, but keeping the ratio $x/\sqrt{t}$ fixed. In this limit,
$\lambda(s)=\sqrt{s^2+2s}\to \sqrt{2s}$ as $s\to 0$.  Then, using the Laplace
inversion
\begin{equation}
\mathcal{L}^{-1}\left[ e^{-a \sqrt{s}}\right]= \frac{a}{\sqrt{4\pi t^3}}\,\,
e^{-a^2/{4t}} 
\label{li.1}
\end{equation}
we find that both $f_{\pm}(x,t)$ converge, in the scaling limit, to the Holtsmark distribution
\begin{equation}
f_{\pm}(x,t) \to \frac{x}{\sqrt{4\pi\, D_0\, t^3}}\, 
e^{-x^2/{4\,D_0\,t}}\quad {\rm where}\quad D_0=\tfrac{1}{2}
\label{fpm.2}
\end{equation}
Now we recall that for a Brownian  particle evolving via
$dx/dt= \sqrt{2D_0}\, \eta(t)$, the first-passage probability $f(x,t)$ is
given precisely~\cite{R01} by the formula in Eq.~\eqref{fpm.2}.  Hence, for
our RTP that evolves via the telegraphic noise in Eq.~\eqref{eqn_Langevin}
with ${\cal D}=0$, its first-passage probability in the scaling limit is
equivalent to that for a normally diffusing particle with diffusion constant
$D_0=1/2$. Indeed, this result is also consistent with our findings in
Eq.~\eqref{xta}, where we showed that, for ${\cal D}=0$,
$\langle x^2(t)\rangle \to t$, which also corresponds to an effective normal
diffusion at late times, with diffusion constant $D_0=1/2$.

To find the behavior of $f_{\pm}(x,t)$ for finite $t$, we need to invert the
Laplace transforms in Eqs.~\eqref{fpm1} exactly.  Fortunately, this can be
done using the following Laplace inversions
  \begin{eqnarray}
\mathcal{L}^{-1} \left(\frac{e^{-x \lambda(s)}}{\lambda(s)[s+1+\lambda(s)]}\right)
    = e^{-t}\frac{\sqrt{t-x}}{\sqrt{t+x}}~
I_1 \big(\sqrt{t^2-x^2}\big)\,,     \nonumber \\
\mathcal{L}^{-1}\left(\frac{e^{-x\lambda(s)}}{\lambda(s)}\right) = 
e^{- t} \,I_0 \big(\sqrt{t^2-x^2}\big)\, , \nonumber  
\end{eqnarray}
where $I_{0,1}(t)$ are modified Bessel functions and 
$\lambda(s)=\sqrt{s^2+2s}$.
Taking the derivative with respect to $x$, we obtain
\begin{align}
  \label{fptsol}
\begin{split}
f_+(x,t)&=\frac{e^{-t}}{t+x}\left[ x ~I_0\big(  \sqrt{t^2-x^2} \big) +  \frac{\sqrt{t-x}}{\sqrt{t+x}}~I_1\big(  \sqrt{t^2-x^2} \big)\right]\theta(t-x)\,,  \\
f_-(x,t)&=e^{- t} \frac{ x}{\sqrt{t^2-x^2}} 
\,I_1 \big(  \sqrt{t^2-x^2} \big)~ \theta(t-x)\,+  e^{- t} \delta(t-x)\,.
\end{split}
\end{align}
These results match  with those obtained by Orsingher \cite{orsingher} using a different approach.

\begin{figure}
\centering
 \includegraphics[width=0.45\textwidth]{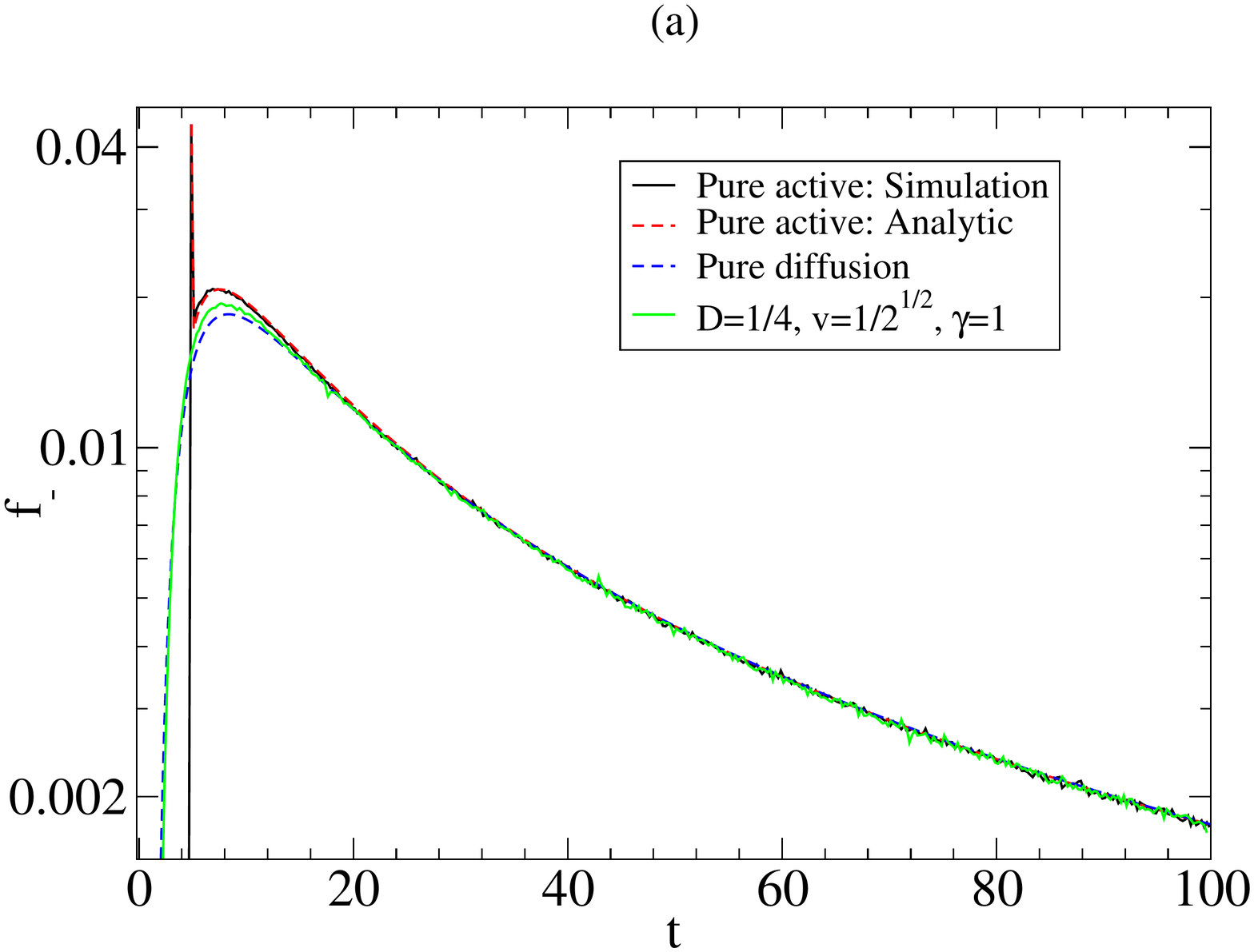} 
 \includegraphics[width=0.45\textwidth]{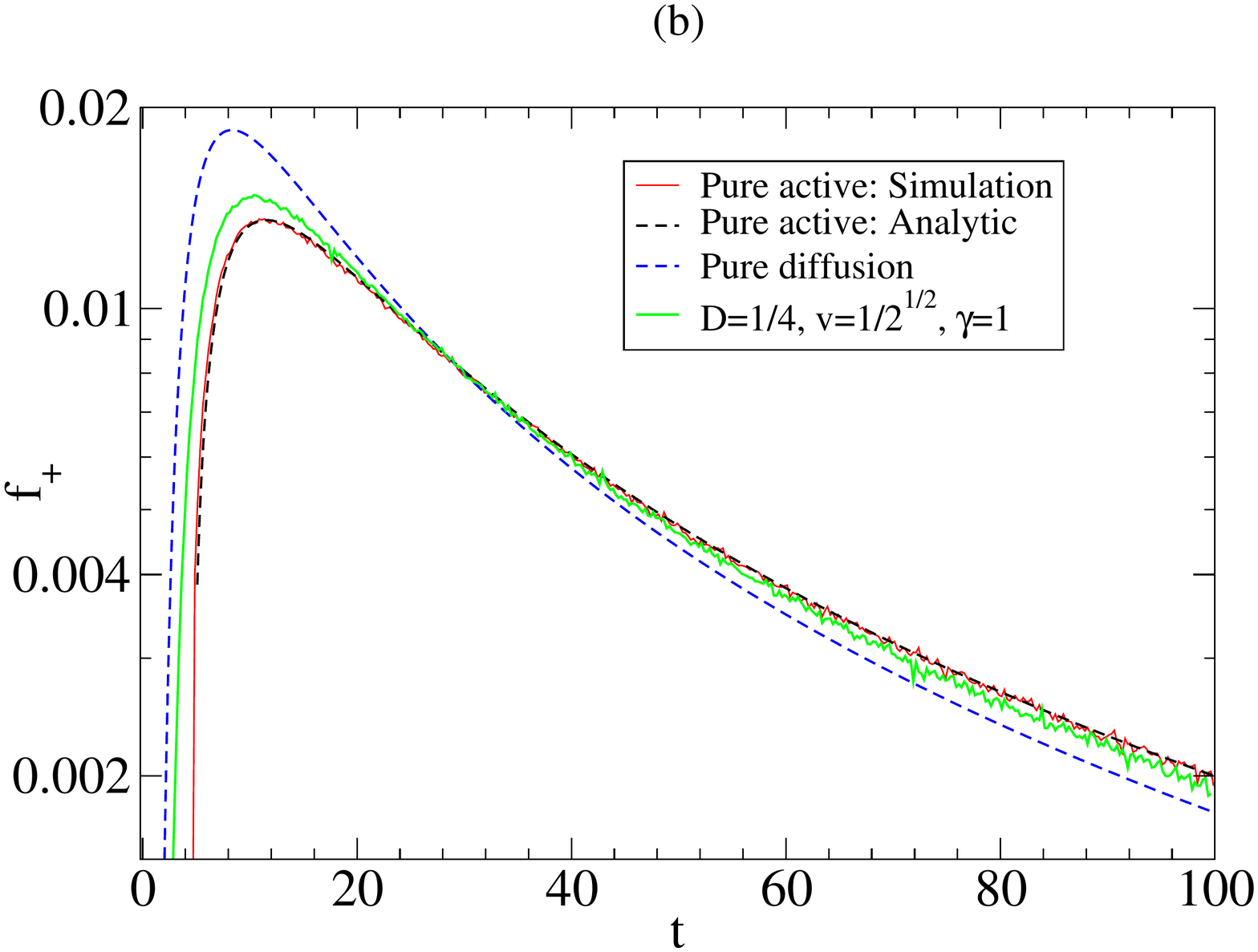} 
 \caption{Comparison of the first-passage probability distributions
   $\gamma f_{\pm}(x\gamma/v,t\gamma)$ from the exact results in (\ref{fptsol}) with with direct
   simulations of (\ref{eqn_Langevin}), with $D \equiv v^2 \mathcal{D} / (2\gamma)=0, v=1, \gamma=1$ (pure active process). The
   starting point is taken to be $x=5$. The colored points correspond to
   simulation results while the black solid lines correspond to the exact
   result. Note that $\gamma f_{-}(x\gamma/v,t\gamma)$ has a $\delta$-function peak at $t=x$
   corresponding to particles which reach the origin without any scattering. For comparison we also plot results for the pure diffusion case (with $D=1/2, v=0, \gamma=0$) and a mixed case. For the mixed case, the parameters are chosen as $D=1/4, v=1/2^{1/2}, \gamma=1$ so that the asymptotic effective diffusion constant is still $D+v^2/(2 \gamma)=1/2$ }
\label{fig:fpd}
\end{figure}

In Fig.~\ref{fig:fpd} we verify these results for the first-passage time distributions with simulations. 
For any given $x$,  the large $t$ limit of \eref{fptsol} can be taken by  using the asymptotic
behavior $I_1(z)\to e^z/\sqrt{2\pi z}$ as $z\to \infty$. This yields, as $t\to\infty$,  for any $x$, 
\begin{align}
f_-(x,t)\simeq  \frac{1}{\sqrt{2\pi}}\frac{x}{t^{3/2}}\, \quad\text{and}\quad
f_+(x,t)\simeq \frac{1}{\sqrt{2\pi}}\frac{x+1}{t^{3/2}}\,.
\label{asym-f}
\end{align}
While the tail of  $f_-(x,t)$ behaves exactly as in the case of Brownian diffusion with a diffusion coefficient $D_0=1/2$ with a starting position $x$,  the tail of $f_+(x,t)$ is equivalent to that in a Brownian diffusion with a starting position $x+1$.  The extra length  $1 \, (=v/\gamma)$  is the average distance the RTP with a positive velocity moves before taking the first turn. In Fig.~\ref{fig:fpd} we compare the asymptotic results  of \eref{asym-f} with numerical simulations and find very good agreement. From \eref{asym-f}, the large time behavior of the survival probabilities are given by
\begin{equation}
S_-(x,t) \sim \frac{x}{\sqrt{t}}\quad\text{and}\quad S_+(x,t)\sim \frac{x+1}{\sqrt{t}}.
\end{equation}
In comparison with a particle with the negative starting velocity, a particle with the positive starting velocity has a higher probability of survival.

\subsubsection{The case ${\cal D}\ne 0$.}

As explained in the beginning of this subsection, the survival probabilities
$S_{\pm}(x,t)$ satisfy the boundary conditions in Eq. (\ref{caldnonzero}),
i.e., in the Laplace domain, $\tilde{S}_{\pm}(x=0,s)=0$. In this case, the
shifted functions $U_{\pm}(x,s)$ each satisfy a fourth-order ordinary
differential equation with constant coefficients, as seen from
Eqs.~\eqref{upm2}. Trying again a solution of the form:
$U_{\pm}(x,s)\sim e^{-\lambda x}$, we find that $\lambda$ has now 4
possible values that are the roots of the fourth-order polynomial
\begin{equation}
\left[{\cal D}\lambda^2-\lambda-(1+s)\right]\left[{\cal 
D}\lambda^2+\lambda-(1+s)\right]=1\, .
\label{poly4}
\end{equation} 
There are $4$ solutions given by $\pm \lambda_1(s)$ and 
$\pm \lambda_2(s)$ where
\begin{align}
  \label{l12s}
  \begin{split}
\lambda_1(s) &= \left[\frac{1+{\cal D}(1+s)-\sqrt{1+4{\cal D}(1+s)+4{\cal 
D}^2}}{2{\cal D}^2}\right]^{1/2}\,, \\ 
\lambda_2(s) &=  \left[\frac{1+{\cal D}(1+s)+\sqrt{1+4{\cal 
D}(1+s)+4{\cal   
D}^2}}{2{\cal D}^2}\right]^{1/2}\, .
\end{split}               
\end{align}
Evidently, $\lambda_1(s)<\lambda_2(s)$. Again discarding the negative roots
$-\lambda_1(s)$ and $-\lambda_2(s)$ (since the solution cannot diverge as
$x\to \infty$), the general solutions of Eqs.~\eqref{upm2} can be written as
\begin{align}
  \begin{split}
U_+(x,s)& =  B_1\, e^{-\lambda_1(s)\, x} + B_2\, e^{-\lambda_2(s)\, x} \,, \\
U_{-}(x,s) &= A_1\, e^{-\lambda_1(s)\, x} + A_2\, e^{-\lambda_2(s)\, x} \,,
\end{split}
\end{align}
where $\lambda_{1,2}(s)$ are given in Eqs.~\eqref{l12s}.  However, these
solutions must also satisfy the individual second-order equations
\eqref{upm1}. This indicates that $A_1$, $A_2$ are related to $B_1$ and
$B_2$. Indeed, by substituting these solutions in Eqs.~\eqref{upm1} gives the
following relations
\begin{align}
  \label{ba12}
  \begin{split}
B_1 &=  -\frac{A_1}{{\cal D}\lambda_1^2-\lambda_1-(1+s)}= -A_1 
\left({\cal D} 
\lambda_1^2+\lambda_1-(1+s)\right) \,, \\
B_2 &= -\frac{A_2}{{\cal D}\lambda_2^2-\lambda_2-(1+s)}= -A_2
\left({\cal D}
      \lambda_2^2+\lambda_2-(1+s)\right)\,.
\end{split}      
\end{align}
Note that we have used Eq.~(\ref{poly4}) to obtain the last two relations.

Hence, the solutions for the survival probabilities are given by
\begin{align}
\label{survpm1}
\begin{split}
\tilde{S}_+(x,s) &=  \frac{1}{s} +B_1\, e^{-\lambda_1\, x}+ B_2\, 
e^{-\lambda_2\, x}\,, \\
\tilde{S}_-(x,s) &=  \frac{1}{s}+ A_1\, e^{-\lambda_1\, x} + A_2\, 
e^{-\lambda_2\, x}\, 
\end{split}
\end{align}
where $B_1$, $B_2$ are related to $A_1$ and $A_2$ via Eqs.~\eqref{ba12}.  We
are still left with two unknown constants $A_1$ and $A_2$. To fix them, we
use the two boundary conditions: $\tilde{S}_{\pm}(x=0, s)=0$. This gives two
linear equations for $A_1$ and $A_2$ whose solution is
\begin{align}
  \label{a12}
  \begin{split}
A_1 & = -\frac{1}{s} \frac{{\cal D}\lambda_2^2 + \lambda_2 -s}{[{\cal D} 
(\lambda_2^2-\lambda_1^2)+ \lambda_2-\lambda_1]}\,,\\
A_2 &=  \frac{1}{s} \frac{{\cal D}\lambda_1^2 + \lambda_1 -s}{[{\cal D}
         (\lambda_2^2-\lambda_1^2)+ \lambda_2-\lambda_1]}\, .
\end{split}         
\end{align}
This then uniquely determines the solutions for the survival probabilities
$\tilde{S}_{\pm}(x,s)$. The corresponding first-passage probabilities are
given by
\begin{align}
\label{fpml2}
\begin{split}
\tilde{f}_{+}(x,s)&= 1-s \tilde{S}_{+}(x,s)= -s\left[ B_1\, 
e^{-\lambda_1\, x} + B_2\, e^{-\lambda_2\, x}\right] \,, \\
\tilde{f}_{-}(x,s)&= 1-s \tilde{S}_{-}(x,s)= -s\left[ A_1\,   
  e^{-\lambda_1\, x} + A_2\, e^{-\lambda_2\, x}\right] \,,
\end{split}
\end{align}
where the constants $A_1$, $A_2$, $B_1$ and $B_2$ are
determined explicitly above and $\lambda_{1,2}(s)$ are given in
Eqs.~\eqref{l12s}.

The first nontrivial check is the limit ${\cal D}\to 0$. In this limit, it is easy to verify, from 
Eqs.~\eqref{l12s}, that 
\begin{equation}
\lambda_1(s) \to \lambda(s)=\sqrt{s^2+2s},\qquad {\rm and} \qquad 
\lambda_2(s)\to \frac{1}{{\cal D}}\to \infty\, .
\label{lD0}
\end{equation}
In addition, one finds that as ${\cal D}\to 0$,
\begin{equation}
A_1\to - \frac{1}{s}, \qquad {\rm and} \qquad A_2\to 0\, ,
\label{a1a2D0}
\end{equation}
and consequently
\begin{equation}
B_1\to -\frac{1}{s (1+s+ \sqrt{s^2+2s})}, \qquad {\rm and} \qquad B_2 \to 
0\, .
\label{b1b2D0}
\end{equation}
We therefore recover the ${\cal D}=0$ results in Eqs.~\eqref{fpm1}.

We now turn to the long-time asymptotic solutions of Eqs.~\eqref{fpml2} for
arbitrary ${\cal D}$.  Hence we consider the $s\to 0$ limit, with finite
${\cal D}$. In this limit, it is easy to check that, to leading order for
small $s$
\begin{equation}
\lambda_1(s)\to \left[\frac{2}{1+2{\cal D}}\, s\right]^{1/2}, \qquad {\rm 
and}\qquad \lambda_2(s) \to \frac{\sqrt{1+2{\cal D}}}{{\cal D}}\, .
\label{l1l2s0}
\end{equation}
Similarly, one can check that to leading order for small $s$
\begin{equation*}
s\, A_1 \to -1, \qquad {\rm and}\qquad s\,A_2\to O(\sqrt{s})\,,
\label{a1a2s0}
\end{equation*}
and consequently 
\begin{equation*}
s\, B_1 \to -1 \qquad {\rm and}\qquad s\,B_2 \to O(\sqrt{s})\, .
\label{b1b2s0}
\end{equation*}
Substituting these results together in Eqs.~\eqref{fpml2}, we find that in
the scaling limit ($s\to 0$, $x\to \infty$ with the product $x\sqrt{s}$
fixed)
\begin{equation}
\tilde{f}_{\pm}(x,s) \to \exp\left[- \sqrt{\frac{2s}{1+2{\cal D}}}\,\,
x\right]\, .
\label{fpmx.1}
\end{equation}
Upon inverting the Laplace transform using Eq.~\eqref{li.1}, we obtain our
final results in the scaling limit
\begin{equation}
f_{\pm}(x,t) \to \frac{x}{\sqrt{4\pi\, D_1\, t^3}}\,\,
e^{-x^2/{4\,D_1\,t}}\,, \quad {\rm where} \quad D_1= {\cal D}+\frac{1}{2}\, .
\label{fpmx.2}
\end{equation}
This result is precisely the same as the first-passage time density of an
ordinary Brownian motion with diffusion constant $D_1= {\cal D} +1/2$. Note
that for $D_1\to D_0=1/2$ as ${\cal D}\to 0$. Moreover, this effective
diffusion constant $D_1= {\cal D}+1/2$ is consistent with our result
$\langle x^2(t)\rangle \to (2{\cal D}+1)\,t$ in Eq.~\eqref{xta}.

Unfortunately, unlike in the ${\cal D}=0$ case, for nonzero ${\cal D}$, we
are not able to obtain the finite time result for $f_{\pm}(x,t)$ explicitly,
due to the fact that the Laplace transforms are difficult to invert.

\subsection{Exit probabilities and exit times in the finite interval}
\label{sec:exit_times}

We now investigate an RTP on a finite interval and address two questions: (a)
the probability for the particle, which starts at $x$, to eventually reach
either of the boundaries, and (b) the mean time for the particle to exit the
interval by either of the boundaries.  Let $E_+(x)$ ($E_-(x)$) denote the
exit probabilities, namely the probability for a particle that starts at $x$
with velocity $+1$ ($-1$) exits through the boundary at $x=-\ell$.  By
comparison, the exit probability to $x=-\ell$ for isotropic diffusion,
$E(x)$, is simply $\frac{1}{2}(1-\frac{x}{\ell})$; that is, the exit
probability decreases linearly with the initial distance from the left edge.

It is easily seen that these hitting probabilities obey the backward
equations~\cite{R01}
\begin{align}
\begin{split}
& \mathcal{D}\, \partial^2_{x}E_+  +  \partial_x E_+ - (E_+-E_-) = 0 \,, \\
& \mathcal{D}\, \partial^2_x E_- - \partial_x E_- + (E_+-E_-) = 0 \,,
\label{E+-}
\end{split}
\end{align}
subject to the appropriate boundary conditions, which are $E_\pm(-\ell)=1$
and $E_\pm(\ell)=0$.  These boundary conditions fix the constants in $E_\pm$
and thus the problem is formally solved.  The calculation is conceptually
straightforward but tedious, and the details were performed by
Mathematica.  The basic steps and the final expressions for the exit
probabilities are given by \eqref{E+-finite} in \ref{app-E-finite}.
\begin{figure}[ht]
\centering
\includegraphics[width=\textwidth]{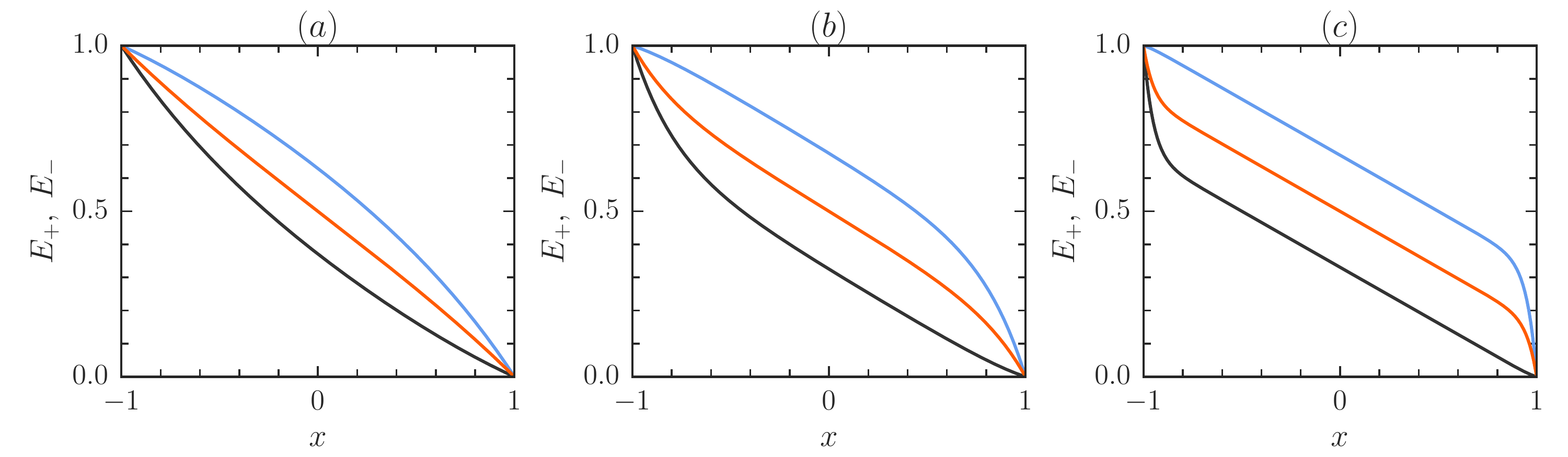}
\caption{The exit probabilities $E_+(x)$ (black) and $E_-(x)$ (blue), and
    their average (orange), as a function of $x$ on an interval of scaled
    length $\ell=1$ for a particle with various diffusion (scaled)
    coefficient (a) $\mathcal{D}=1.0$, (b) $\mathcal{D}=0.25$, and (c)
    $\mathcal{D}=0.05$.}
\label{fig:E+-}
\end{figure}

Fig.~\ref{fig:E+-} shows the exit probabilities $E_\pm(x)$ for representative
values of the dimensionless diffusion coefficient $\mathcal{D}$.  As one
expects, for $\mathcal{D}\gg 1$, the exit probabilities are close to the
isotropic random-walk form $\frac{1}{2}(1-\frac{x}{\ell})$.  However, for
$\mathcal{D}\ll 1$, $E_+$ and $E_-$ become very distinct.  Moreover, the exit
probability $E_+$ decreases much more rapidly with $x$ than
$(1-\frac{x}{\ell})/2$, while $E_-$ decreases much more slowly.  Notice also
that
$\mathcal{E}(x)\equiv \frac{1}{2}[E_+(x)+E_-(x)]\ne
\frac{1}{2}(1-\frac{x}{\ell})$.  That is, the exit probability, averaged over
the two velocity states deviates significantly from the corresponding exit
probability for unbiased diffusion.

Let us now turn to the exit times.  Let $t_+(x,t)$ ($t_-(x,t)$) be the mean
first-passage time (to either boundary) for a particle that is at $x$ and is
also in the $+$ ($-$) state.  Again using the formalism given in~\cite{R01},
it is easily seen that these exit times obey the backward equations
\begin{align}
\begin{split}
\label{t+-}
& \mathcal{D} \partial^2_{x}t_+  +  \partial_x t_+ - (t_+-t_-) = -1 \,, \\
& \mathcal{D} \partial^2_x t_- - \partial_x t_- + (t_+-t_-) = -1 \,,
\end{split}
\end{align}
with boundary conditions $t_\pm(\pm\ell)=0$, which corresponds to the
particle being immediately absorbed if it starts at either end of the
interval.
\begin{figure}[t]
\centering
\includegraphics[width=\textwidth]{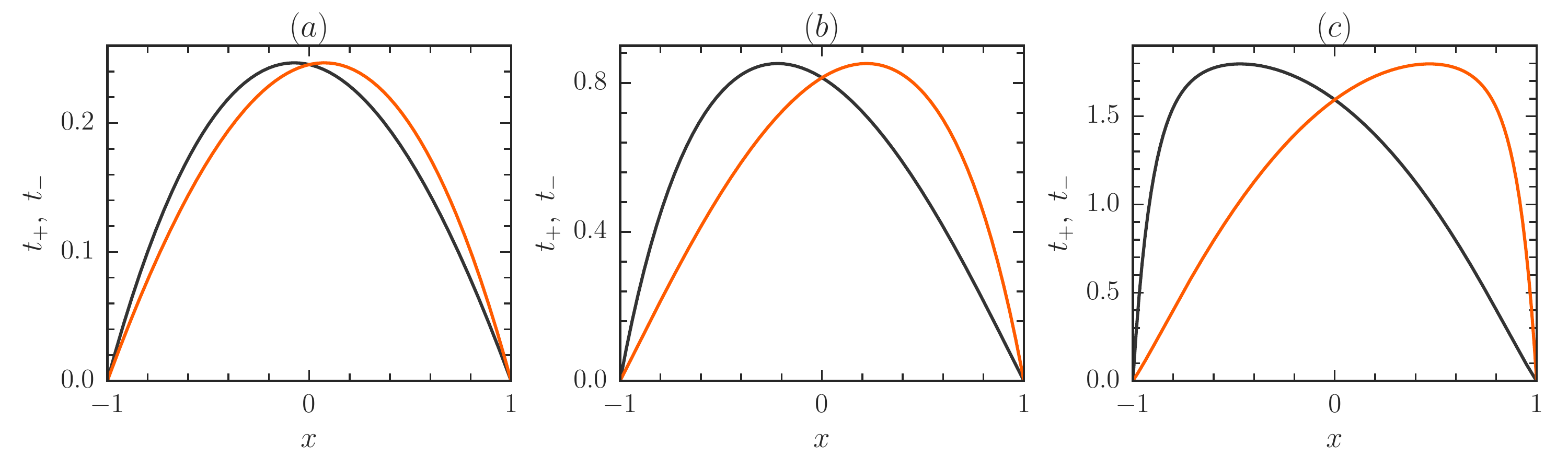}
\caption{{The unconditional exit times $t_+(x)$ (black) and $t_-(x)$ (orange) to
  either side of the interval as a function of $x$ for an interval of scaled length
  $\ell=1$ for a particle with various diffusion (scaled) coefficient (a) $\mathcal{D}=1.0$, (b) $\mathcal{D}=0.5$, and (c)
  $\mathcal{D}=0.1$.}}
\label{fig:t+-}
\end{figure}
The solution to Eqs.~\eqref{t+-} are obtained using the same approach as that
given for Eqs.~\eqref{E+-} (see \ref{app-t}).  While the resulting
expressions for $t_\pm$ for the finite interval are too long to be displayed,
the form of the first-passage times are easily visualized
(Fig.~\ref{fig:t+-}).  For small bias velocity the scaled diffusion constant
$\mathcal{D}=D\gamma/v^2$ (see Sec.~\ref{sec:model}) is large. In this limit
the diffusive part of the dynamics for both types of the particles
dominates. As a result, $t_+\approx t_-$, and both $t_+$ and $t_-$ become very close
to the exit probability for isotropic diffusion
$t=(\ell^2-x^2)/2\mathcal{D}$.  On the other hand, if $v$ is increased
$\mathcal{D}$ is decreases and the active contribution to the motion
dominates. In this limit the exit times $t_+$ and $t_-$ strongly deviate from
each other.

\section{Summary}
\label{sec:summary}
We studied a one-dimensional model of run-and-tumble particles in the
presence of an additional diffusion term. On the infinite line we find that
an initial localized distribution of particles evolves to a Gaussian
distribution at long times, with the diffusion constant renormalized by the
active particle speed and the tumble rate, while at intermediate times the
density distribution can have a multimodal structure.

In a finite domain with reflecting walls, we found that the RTPs reach a
steady-state with peaks in the density distributions at the boundaries, which
is in agreement with earlier observations of particle accumulation at
walls. We also studied the approach to the steady-state by examining the
spectral structure of the corresponding Fokker-Planck operator.  The
eigenvalues of this operator appear as the zeros of a complicated
determinant, and finding them is highly non-trivial, even numerically. We
numerically evaluated the two eigenvalues with largest real parts.  It is an
interesting mathematical problem to find the full spectrum as well as the
associated eigenvectors of the Fokker-Planck operator.

We also investigated the first-passage probability distribution of an RTP on
the semi-infinite line and obtained an explicit closed form expression for
the distribution in the limiting case of zero diffusion and in the more
challenging case of non-zero diffusion.  In a finite domain, we obtained
exact results for the exit time probability and the mean exit time.

We believe that our results for non-interacting RTPs in one dimension will be
informative for the study of models of other active particle systems in
higher dimensions.  Another possible extension of this work is to study
active particles in external potentials and in the presence of mutual
interactions.  It will be interesting to verify some of our analytic
observations in experimental systems such as vibrated granular systems and
Janus particles~\cite{bechinger_active_2016}.

\section{Acknowledgment}
KM acknowledges the S. N. Bhatt Memorial Excellence Fellowship Program 2016 at ICTS, and INSPIRE-SHE (awarded by DST, Government of India) for funding her research. VJ is supported by a post-doctoral fellowship in the Max Planck partner group at ICTS 
 AD, AK, SNM and SS acknowledge support from the Indo-French Centre for the promotion of advanced research (IFCPAR) under Project No. 5604-2. 
 AD, AK, SNM and SS also acknowledge the large deviation theory program at ICTS (code: ICTS/Prog-ldt/2017/8) during which many discussions were held. 
 SNM wishes to thank U. Basu. M.R. Evans, A. Rosso,
and G. Schehr for useful discussions, and acknowledges a Simon foundation
grant from ICTS.  
 KVK's research is supported by the Department of Biotechnology, India, through a Ramalingaswami reentry fellowship and by the Max Planck Society and the Department of Science and Technology, India, through a Max Planck Partner Group at ICTS-TIFR.
SR acknowledges support from grants DMR16-08211 and DMR-1623243 from the National Science Foundation and from the ICTS for supporting his participation in the Bangalore school on statistical physics - VIII (code: ICTS/Prog-bssp/2017/06).  He also thanks Uttam Bhat for many helpful discussions.

\section*{References}

\providecommand{\newblock}{}

\appendix

\section{Steady-State Probability Distribution in the Interval}
\label{app:interval}

We define $Q=P_+-P_-$ and use $P= P_{+}+P_{-}$.  With these definitions,
(\ref{eqn:Pss}) can be rewritten as:
\begin{align}
  \label{steadystate}
  \begin{split}
\mathcal{D} \partial^2_{x} P- \partial_x Q   =0\,, \\
\mathcal{D} \partial^2_{x} Q- \partial_x P -2  Q=0\,,
  \end{split}
\end{align}
and the boundary conditions (\ref{BC12})  now read 
\begin{align}
  \label{boundary}
  \begin{split}
( \mathcal{D}\partial_x Q -P)|_{x=\pm \ell}=0\,,  \\
  ( \mathcal{D}\partial_x P -Q)|_{x=\pm \ell}=0\,.
  \end{split}
\end{align}
Integrating the first of (\ref{steadystate}), we get
$\mathcal{D} \partial_{x}P-Q+C_1=0 $ where $C_1$ is an integration
constant. From the second boundary condition in (\ref{boundary}), we get
$C_1=0$. Hence $\mathcal{D} \partial_{x}P(x)=Q(x)$ for all
$x \in [-\ell,\ell]$, and substituting this into the second of
(\ref{steadystate}) leads to
\begin{align}
\mathcal{D} \partial^2_{x} Q -\left( 2+\mathcal{D}^{-1}  \right) Q=0\,.
\label{eqnQ-2nd}
\end{align}
This equation has the general solution
\begin{align}
 Q= a e^{\mu x} +b e^{-\mu x},~~~\textrm{with}~~\mu=\frac{\sqrt{2\mathcal{D}+1}}{\mathcal{D}},
 \label{QSoln}
\end{align}
and $a$ and $b$ are constants to be determined.

Once $Q(x)$ is known, $P(x)$ can be obtained by integrating $\mathcal{D} \partial_{x}P(x,t)=Q(x,t)$:
\begin{align}
P(x)= \frac{1}{\mathcal{D} \mu}\bigg( a e^{\mu x} -  b e^{-\mu x}\bigg)+C_2,
\label{PSoln}
\end{align}
where $C_2$ is another integration constant. The three constants $a, b$ and
$C_2$ can be obtained using the boundary conditions and the normalization
condition. Substituting the solutions (\ref{QSoln}) and (\ref{PSoln}) into
the first of (\ref{boundary}) gives
\begin{align}
\bigg( a e^{\mu \ell} -  b e^{-\mu \ell}\bigg) =\frac{\mu}{2 }  C_2 \,,~~~\textrm{and}~~~
\bigg( a e^{-\mu \ell} -  b e^{\mu \ell}\bigg)=\frac{\mu}{2 } C_2 \,,
\end{align}
whose solution is
\begin{align}
 a=\frac{\mu}{4 \cosh(\mu \ell) }C_2; \qquad \textrm{and} \quad b=-a.
 \label{A_fixing}
\end{align}

Finally, we invoke the normalization condition $\int_{-\ell}^{\ell}P(x) dx=1$
to obtain
\begin{align}
\fl \qquad ~~ Q(x)= \frac{ (2\mathcal{D}+1)}{2\mathcal{D}\left[ \sinh\left(\frac{\sqrt{2\mathcal{D}+1} }{\mathcal{D}} \ell\right)+2\ell \sqrt{2\mathcal{D}+1} \cosh\left(\frac{\sqrt{2\mathcal{D}+1} }{\mathcal{D}} \ell\right)\right]} \sinh\left(\frac{\sqrt{2\mathcal{D}+1} }{\mathcal{D}} x\right)\,, \label{expression_Q}
\end{align}
and
\begin{align}
P(x)={\left[\frac{\tanh\left(\frac{\sqrt{2\mathcal{D}+1} }{\mathcal{D}} \ell\right)}{\sqrt{2\mathcal{D}+1}} +2\ell\right]}^{-1}\left[\frac{\cosh\left(\frac{\sqrt{2\mathcal{D}+1}}{\mathcal{D}} x\right)}{2 \mathcal{D} \cosh\left(\frac{\sqrt{2\mathcal{D}+1}}{\mathcal{D}} \ell \right)}  +1\right]\,.
\label{expression_P-app}
\end{align}
The latter is \eqref{expression_P} in the main text.

\section{Time-Dependent Probability Distribution in the Interval}
\label{app:interval-td}

We now construct the  eigenstates $\phi_n^\pm(x)$. First we try a solution of the form
\begin{eqnarray}
  \begin{pmatrix} \phi^{+}(x) \\  \phi^{-}(x) \end{pmatrix}
  =  e^{\beta x} \begin{pmatrix} {r}^{+} \\  {r}^{-} \end{pmatrix}\,.
\label{solutin_Pbar+Pbar-}
\end{eqnarray}
Inserting this form in \eqref{eqn:lambda}, we get 
\begin{eqnarray}
\begin{pmatrix} \mathcal{D}~\beta^2 - \beta -\lambda-1& 1 \\ 1 & \mathcal{D}~\beta^2 +\beta-\lambda-1 \end{pmatrix}
\begin{pmatrix} r^{+} \\  r^{-}  \end{pmatrix} =  \begin{pmatrix} 0\\0 \end{pmatrix}\,.
\label{eigen_value_eqn}
\end{eqnarray}
To get non-zero solutions for $\bar{r}_\pm$, we require the determinant of the matrix in the above equation to
be zero: $(\mathcal{D}~\beta^2 -\lambda-1)^2-(\beta^2+1)=0$ which provides
$\beta$ as function of $\lambda$.  This is a fourth order equation in $\beta$
whose solutions are
\begin{align}
\beta_{\sigma\tau}(\lambda)&=\sigma \sqrt{\frac{2\mathcal{D} +1+ 2\mathcal{D}\lambda +\tau \sqrt{(2\mathcal{D} +1)^2+ 4\mathcal{D}\lambda}}{2\mathcal{D}^2}}\,,\label{beta-st}
\end{align}
where $\sigma=\pm 1$ and $\tau=\pm1$.  Corresponding to the resulting four
values of $\beta$, we get the four corresponding solutions 
\begin{align}
\begin{pmatrix} r^{+}_{\sigma \tau} \\  r^{-}_{\sigma \tau}  \end{pmatrix}
 =& \begin{pmatrix} 1 \\ \alpha_{\sigma \tau}   \end{pmatrix}\,, 
\end{align}
where $\alpha_{\sigma \tau}=  -{(\mathcal{D}\beta_{\sigma \tau}^2
  -\beta_{\sigma \tau} -\lambda-1 )}$ and $\sigma= \tau=\pm1$.
We use these four states to construct the eigenstates
$[\phi^+_n(x),\phi^-_n(x)]$ that satisfy the boundary conditions. Thus let
\begin{equation}
\begin{pmatrix} \phi^+_n(x,t) \\ \phi^-_n(x,t) \end{pmatrix}=
\sum_{\sigma=\pm1} \sum_{\tau=\pm1} C^{\sigma \tau}_n~e^{\beta_{\sigma \tau}^{(n)} x} \begin{pmatrix} 1\\ \alpha_{\sigma \tau} \end{pmatrix} \,. \label{phisoln}
\end{equation}
Substituting this solution into the required boundary conditions (\ref{BC12}), and after some rearrangement, we get
\begin{align}
  {M}  \begin{pmatrix} C^{++}_n \\ C^{+-}_n \\ C^{-+}_n \\ C^{--}_n\end{pmatrix} =
  0~,~~~~ 
{\rm where}~~~
  {M} = 
\begin{pmatrix}
\nu_{++}^{+-}    &       \nu_{+-}^{+-}    &    \nu_{-+}^{+-}   &   \nu_{--}^{+-}   \\  
\nu_{++}^{++}  &       \nu_{+-}^{++}     &    \nu_{-+}^{++}   &   \nu_{--}^{++}   \\  
\nu_{++}^{--}    &       \nu_{+-}^{--}     &    \nu_{-+}^{--}   &   \nu_{--}^{--}   \\  
\nu_{++}^{-+}    &       \nu_{+-}^{-+}     &    \nu_{-+}^{-+}   &   \nu_{--}^{-+} \\   
\end{pmatrix}~, \label{c-eq}
 \end{align}  
with $\nu_{\sigma \tau}^{r s}= e^{r \beta_{\sigma \tau} \ell}(\mathcal{D}\beta_{\sigma \tau} +s  )$, and $\sigma,\tau,r,s$ allowed to take values $\pm 1$. 
 To get non-zero solutions for $C^{\sigma\tau}_n$, we require $\det(M)=0$. This 
equation has both real and imaginary parts and both have to be set to zero. 
This is possible only at certain values (in general complex) of $\lambda$ and these values then give us the required eigenvalue set $\{\lambda_n\}$, $n=0,1,2,\ldots$. We assume that the eigenvalues are ordered according to decreasing value of their real part. 
For each allowed $\lambda_n$ one can find the
 corresponding value of $\beta^{n}_{\sigma \tau}$ from (\ref{beta-st}). If the $\beta$s are
 non-degenerate then the associated eigenvector
 $(C^{++}_n,~C^{+-}_n,~C^{-+}_n,~C^{--}_n)^T$ can be obtained from
 \eqref{c-eq}. This then determines the eigenstates completely, up to a
 normalization constant.

We expect  that there should be a real largest eigenvalue $\lambda_0=0$ corresponding to the steady-state and this was already determined, see (\ref{expression_P}).  This  solution can be recovered from our present approach  
but needs some  extra care since  for this case, $\beta_{\pm-}=0$ and
 $\beta_{\pm+}=\pm \sqrt{2\mathcal{D}+1}/\mathcal{D}$.  The two independent states
 corresponding to $\beta=0$  are given by
\begin{eqnarray}
\begin{pmatrix} r^{+}_{+-} \\  r^{-}_{+-}  \end{pmatrix} = \left (\begin{array}{c} 1 \\ 1 \end{array} \right )~,~~
 \begin{pmatrix} r^{+}_{--} \\  r^{-}_{--}  \end{pmatrix}= \left (\begin{array}{c}  x \\ 1+x \end{array} \right )~.
\end{eqnarray}
Taking a linear combination and imposing the boundary conditions leads us to
the solution given in Eqs.~\eqref{expression_P} and \eqref{expression_Q}.  

\section{Solution for $E_\pm(x)$ on the Finite Interval}
\label{app-E-finite}

To solve Eqs.~\eqref{E+-}, we first define $S_e=E_++E_-$ and
$\Delta_e=E_+-E_-$ to recast \eqref{E+-} as
\begin{align}
\begin{split}
\label{SD}
&\mathcal{D}\,S_e''+\Delta_e'=0\,, \\
&\mathcal{D}\,\Delta_e''+S_e'=2\Delta_e\,.
\end{split}
\end{align}
Differentiating the second of \eqref{SD} and using the first to eliminate
$S_e''$ gives $\Delta_e'''-\alpha^2\Delta_e'=0$, with
\begin{equation*}
  \alpha^2 =\frac{1}{\mathcal{D}^2}+ \frac{2}{\mathcal{D}}\,.
\end{equation*}
The solution for $\delta_e\equiv \Delta_e'$ is $\delta_e= Ae^{\alpha x}+Be^{-\alpha
  x}$, where $A$ and $B$ are constants.  Integrating once gives $\Delta_e$ and
integrating $\mathcal{D}S_e''=-\Delta_e'$ gives $S_e$.  The final result is
\begin{align}
\begin{split}
\label{SD-final}
\Delta_e &= \frac{A}{\alpha}\, e^{\alpha x}- \frac{B}{\alpha}\, e^{-\alpha x} +C\,,\\
S_e &=  -\frac{A}{\mathcal{D}\alpha^2}\, e^{\alpha x}- \frac{B}{\mathcal{D}\alpha^2} \,e^{-\alpha x} +Ex+F\,,
\end{split}
\end{align}
where $C,E,F$ are constants.  However, to satisfy the second of
Eqs.~\eqref{SD}, we must have $E=2C$.  Using this and finally solving for
$E_\pm$ gives
\begin{align}
\begin{split}
\label{E+-final}
E_+(x) &=\phantom{-} \tfrac{1}{2}A\, e^{\alpha x}\left(\frac{1}{\alpha}-\frac{1}{\mathcal{D}\alpha^2}\right) 
-\tfrac{1}{2}B\, e^{-\alpha x}\left(\frac{1}{\alpha}+\frac{1}{\mathcal{D}\alpha^2}\right)
+Cx+\tfrac{1}{2}(F+C)\,,\\
E_-(x) &= -\tfrac{1}{2}A\, e^{\alpha x}\left(\frac{1}{\alpha}+\frac{1}{\mathcal{D}\alpha^2}\right) 
+\tfrac{1}{2}B\, e^{-\alpha x}\left(\frac{1}{\alpha}-\frac{1}{\mathcal{D}\alpha^2}\right)
+Cx+\tfrac{1}{2}(F-C)\,.\\
\end{split}
\end{align}

For exit via the left edge of the finite interval $[0,\ell]$, the appropriate
boundary conditions are $E_\pm(0)=1$ and $E_\pm(\ell)=0$.  Thus, from
Eqs.~\eqref{E+-final}, we need to solve
\begin{align}
\begin{split}
\label{Cs}
&A\gamma_--B\gamma_+ +\tfrac{1}{2}(F+C)=1\,,\\
-&A\gamma_++B\gamma_- +\tfrac{1}{2}(F-C)=1\,,\\
&A\,\gamma_-\,e^{\alpha \ell}-B\,\gamma_+\, e^{-\alpha \ell}+C\ell+\tfrac{1}{2}(F+C)=0\,,\\
-&A\,\gamma_+\,e^{\alpha \ell}+B\, \gamma_-\,e^{-\alpha \ell} +C\ell+\tfrac{1}{2}(F-C)=0\,,\\
\end{split}
\end{align}
where we have introduced 
\begin{equation*}
\gamma_\pm= \frac{1}{2}\Big(\frac{1}{\alpha}\pm\frac{1}{\mathcal{D}\alpha^2}\Big).
\end{equation*}

Solving these four linear equations by Mathematica, substituting the
coefficients $A,B,F$, and $C$ into \eqref{E+-final}, and then performing some
simplifications, the exit probabilities are:
\begin{align}
\begin{split}  
\label{E+-finite}
E_+(x)&= \frac{e^{\alpha \ell}\left[(\ell-x)-\alpha\gamma_-\right]
+ \left[(\ell-x)+\alpha\gamma_+\right]  +\gamma_+ e^{\alpha(\ell-x)}+\gamma_- e^{\alpha \ell}}
{e^{\alpha \ell}\left[{\ell}+\frac{1}{\mathcal{D}\alpha}\right]+
\left[{\ell}-\frac{1}{\mathcal{D}\alpha}\right]}\,,\\
E_-(x)&= \frac{e^{\alpha \ell}\left[(\ell-x)+\alpha\gamma_+\right]
+ \left[(\ell-x)+\alpha\gamma_-\right]  -\gamma_- e^{\alpha(\ell-x)}-\gamma_+ e^{\alpha x}}
{e^{\alpha \ell}\left[{\ell}+\frac{1}{\mathcal{D}\alpha}\right]+
\left[{\ell}-\frac{1}{\mathcal{D}\alpha}\right]}\,.\\
\end{split}
\end{align}
Some representative graphs of $E_\pm(x)$ are given in Fig.~\ref{fig:E+-}.

\section{Solution for $t_\pm(x)$ on the Finite Interval }
\label{app-t}

To solve \eqref{t+-}, we again define $S_t=t_++t_-$, $\Delta_t=t_+-t_-$ to give
\begin{align}
\begin{split}
\label{SDt}
&\mathcal{D}S_t''+\Delta_t'=-2\,, \\
&\mathcal{D}\Delta_t''+S_t'=2\Delta_t\,.
\end{split}
\end{align}
We follow similar steps to those in \ref{app-E-finite} to obtain
\begin{align}
\begin{split}
\Delta_t &= \frac{A}{\alpha}\, e^{\alpha x}- \frac{B}{\alpha}\, e^{-\alpha x} 
-\frac{2x}{(\mathcal{D}\alpha)^2}+C\,,\\
S_t &=  -\frac{A}{\mathcal{D}\alpha^2}\, e^{\alpha x}- \frac{B}{\mathcal{D}\alpha^2} \,e^{-\alpha x} -
\left(1-\frac{1}{(\mathcal{D}\alpha)^2}\right)\frac{x^2}{\mathcal{D}}+2Cx+F\,,
\end{split}
\end{align}
where $C$ and $F$ are constants, and the additional terms compared to those in
\eqref{SD-final} stem from the additional inhomogeneous term in
Eq.~\eqref{SDt} compared to \eqref{SD}.  The solutions for $t_\pm$ are
\begin{align}
\begin{split} 
\label{t+--int}
t_+(x)&= \phantom{-}A\gamma_- e^{\alpha x}-B\gamma_+ e^{-\alpha x}
-\left(1-\frac{1}{(\mathcal{D}\alpha)^2}\right)\frac{x^2}{2\mathcal{D}}+Cx
-\frac{x}{(\mathcal{D}\alpha)^2}
+\tfrac{1}{2}(F+C)\,,\\
t_-(x)&=-A\gamma_+ e^{\alpha x}+B\gamma_- e^{-\alpha  x}
-\left(1-\frac{1}{(\mathcal{D}\alpha)^2}\right)\frac{x^2}{2\mathcal{D}}+Cx
+\frac{x}{(\mathcal{D}\alpha)^2}
+\tfrac{1}{2}(F-C)\,.
\end{split}
\end{align}

For the exit times $t_\pm$ in a finite interval of length $\ell$, the
boundary conditions are $t_\pm(0)=t_\pm(\ell)=0$.  Applying these boundary
conditions to \eqref{t+--int} fixes the constants $A,B,C$ and $F$, from which
the results shown in Fig.~\ref{fig:t+-} are obtained, again using
Mathematica.


\begin{thebibliography}{10}
\expandafter\ifx\csname url\endcsname\relax
  \def\url#1{{\tt #1}}\fi
\providecommand{\eprint}[2][]{\url{#2}}

\bibitem{ramaswamy_active_2017}
Ramaswamy S 2017 {\em Journal of Statistical Mechanics: Theory and
  Experiment\/} {\bf 2017} 054002

\bibitem{prost_active_2015}
Prost J, J\"ulicher F and Joanny J~F 2015 {\em Nature Physics\/} {\bf 11}
  111--117

\bibitem{marchetti_hydrodynamics_2013}
Marchetti M~C, Joanny J~F, Ramaswamy S, Liverpool T~B, Prost J, Rao M and Simha
  R~A 2013 {\em Reviews of Modern Physics\/} {\bf 85} 1143--1189

\bibitem{BB72}
Berg H~C and A B~D 1972 {\em Nature\/} {\bf 239} 500

\bibitem{BP77}
Berg H and M P~E 1977 {\em Biophys. J.\/} {\bf 20} 193

\bibitem{DZ88}
Devreotes P~N and Zigmond S~H 1988 {\em Annu. Rev. Cell Biol.\/} {\bf 4} 649

\bibitem{bechinger_active_2016}
Bechinger C, Di~Leonardo R, L{\"o}wen H, Reichhardt C, Volpe G and Volpe G 2016
  {\em Reviews of Modern Physics\/} {\bf 88} 045006

\bibitem{cates_when_2013}
Cates M~E and Tailleur J 2013 {\em {EPL} (Europhysics Letters)\/} {\bf 101}
  20010

  
\bibitem{cates_when_2009}
Tailleur J and   Cates M~E 2009 {\em {EPL} (Europhysics Letters)\/} {\bf 86}
60002

\bibitem{Enculescu_when_2011}
Enculescu M and Stark H 2011 {\em Phys. Rev. Lett.\/} {\bf 107} 058301
  
  
\bibitem{Lee_when_2011}
Lee C~F 2013 {\em New Journal of Physics \/} {\bf 15} 055007  

\bibitem{Fily_when_2014}
Fily Y,  Baskaran A   and  Hagan M~F 2014 {\em Soft Matter \/} {\bf 10} 5609 

\bibitem{Szamel_when_2014}
Szamel  G  2014 {\em Physical Review E \/} {\bf 90} 012111


\bibitem{Solon_when_2015}
Solon A~P, Cates M~E and Tailleur J  2015 {\em EPJST \/} {\bf 224} 1231

\bibitem{Vachier_when_2017}
Vachier J and Mazza M~G  2017 {\em arxiv:1709.07488 \/} 


\bibitem{Hermann_when_2017}
Hermann S and Schmidt M  2017 {\em arxiv:1712.08553 \/} 


\bibitem{romanczuk_active_2012}
Romanczuk P, B\"ar M, Ebeling W, Lindner B and Schimansky-Geier L 2012 {\em The
  European Physical Journal Special Topics\/} {\bf 202} 1--162

\bibitem{evans2016}
Slowman A~B, Evans M~R and Blythe R~A 2016 {\em Phys. Rev. Lett.\/} {\bf
  116}(21) 218101

\bibitem{evans2017}
Slowman A~B, Evans M~R and Blythe R~A 2017 {{\em Journal of Physics A: Mathematical and Theoretical\/} {\bf  50} 375601} 


\bibitem{stenhammar_activity-induced_2015}
Stenhammar J, Wittkowski R, Marenduzzo D and Cates M~E 2015 {\em Physical
  Review Letters\/} {\bf 114} 018301

\bibitem{reichhardt_ratchet_2017}
Reichhardt C~J~O and Reichhardt C 2017 {\em Annual Review of Condensed Matter
  Physics\/} {\bf 8} 51--75

\bibitem{cates_motility-induced_2015}
Cates M~E and Tailleur J 2015 {\em Annual Review of Condensed Matter Physics\/}
  {\bf 6} 219--244

\bibitem{PhysRevLett.110.055701}
Redner G~S, Hagan M~F and Baskaran A 2013 {\em Phys. Rev. Lett.\/} {\bf 110}(5)
  055701

\bibitem{stenhammar_continuum_2013}
Stenhammar J, Tiribocchi A, Allen R~J, Marenduzzo D and Cates M~E 2013 {\em
  Physical Review Letters\/} {\bf 111} 145702

\bibitem{patch_kinetics_2017}
Patch A, Yllanes D and Marchetti M~C 2017 {\em Physical Review E\/} {\bf 95}
  012601

\bibitem{solon_pressure_2015_1}
Solon A~P, Fily Y, Baskaran A, Cates M~E, Kafri Y, Kardar M and Tailleur J 2015
  {\em Nature Physics\/} {\bf 11} 673--678

\bibitem{solon_pressure_2015_2}
Solon A~P, Stenhammar J, Wittkowski R, Kardar M, Kafri Y, Cates M~E and
  Tailleur J 2015 {\em Physical Review Letters\/} {\bf 114} 198301

\bibitem{junot_active_2017}
Junot G, Briand G, Ledesma-Alonso R and Dauchot O 2017 {\em Physical Review
  Letters\/} {\bf 119} 028002

\bibitem{fox1986uniform}
Fox R~F 1986 {\em Physical Review A\/} {\bf 34} 4525

\bibitem{hanggi1995colored}
H{\"a}nggi P and Jung P 1995 {\em Advances in chemical physics\/} {\bf 89}
  239--326

\bibitem{dhar_triple_2002}
Dhar A and Chaudhuri D 2002 {\em Physical Review Letters\/} {\bf 89} 065502

\bibitem{supurna_2002}
Samuel J  and Sinha S   2002 {\em Physical Review E\/} {\bf 66} 050801

\bibitem{kurzthaler_intermediate_2016}
Kurzthaler C, Leitmann S and Franosch T 2016 {\em Scientific Reports\/} {\bf 6}
  36702

\bibitem{Tailleur2008}
Tailleur J and Cates M 2008 {\em Physical Review Letters\/} {\bf 100} 218103

\bibitem{argun_non-boltzmann_2016}
Argun A, Moradi A~R, Pince E, Bagci G~B, Imparato A and Volpe G 2016 {\em
  Physical Review E\/} {\bf 94} 062150

\bibitem{das2017confined}
Das S, Gompper G and Winkler R G 2017, { \em New Journal of Physics}

\bibitem{klein_spontaneous_2016}
Klein S, Appert-Rolland C and Evans M~R 2016 {\em Journal of Statistical
  Mechanics: Theory and Experiment\/} {\bf 2016} 093206

\bibitem{maggi_multidimensional_2015}
Maggi C, Marconi U~M~B, Gnan N and Leonardo R~D 2015 {\em Scientific Reports\/}
  {\bf 5} 10742

\bibitem{wagner_steady-state_2017}
Wagner C~G, Hagan M~F and Baskaran A 2017 {\em Journal of Statistical
  Mechanics: Theory and Experiment\/} {\bf 2017} 043203

\bibitem{angelani_confined_2017}
Angelani L 2017 {\em Journal of Physics A: Mathematical and Theoretical\/} {\bf
  50} 325601

\bibitem{elgeti2015run}
Elgeti J and Gompper G 2015 {\em EPL (Europhysics Letters)\/} {\bf 109} 58003

\bibitem{Taylor186}
Taylor G~I 1953 {\em Proceedings of the Royal Society of London A:
  Mathematical, Physical and Engineering Sciences\/} {\bf 219} 186--203

\bibitem{Aris67}
Aris R 1956 {\em Proceedings of the Royal Society of London A: Mathematical,
  Physical and Engineering Sciences\/} {\bf 235} 67--77

\bibitem{PhysRevLett.57.996}
de~Arcangelis L, Koplik J, Redner S and Wilkinson D 1986 {\em Phys. Rev.
  Lett.\/} {\bf 57}(8) 996--999

\bibitem{PhysRevA.37.2619}
Koplik J, Redner S and Wilkinson D 1988 {\em Phys. Rev. A\/} {\bf 37}(7) 2619--2636

\bibitem{whittaker1910history}
Whittaker E~T 1910 {\em A History of the Theories of Aether and Electricity
  from the Age of Descartes to the Close of the Nineteenth Century\/}
  (Longmans, Green and Company)

\bibitem{weiss2002some}
Weiss G~H 2002 {\em Physica A: Statistical Mechanics and its Applications\/}
  {\bf 311} 381--410

\bibitem{PhysRevA.45.7207}
Ben-Naim E, Redner S and ben Avraham D 1992 {\em Phys. Rev. A\/} {\bf 45}(10)
  7207--7213

\bibitem{bricard_2013}
{Antoine Bricard}, {Jean-Baptiste Caussin}, {Nicolas Desreumaux}, {Olivier
  Dauchot} and {Denis Bartolo} 2013 {\em Nature\/} {\bf 503} 95--98

\bibitem{di_leonardo_bacterial_2010}
Di~Leonardo R, Angelani L, Dell'Arciprete D, Ruocco G, Iebba V, Schippa S,
  Conte M~P, Mecarini F, De~Angelis F and Di~Fabrizio E 2010 {\em Proceedings
  of the National Academy of Sciences\/} {\bf 107} 9541--9545

\bibitem{R01}
Redner S 2001 {\em A Guide to First-Passage Processes\/} (Cambridge University Press)
  
\bibitem{review_persistence}
Bray A~J., Majumdar S~N., and Schehr, G 2013 {\em Adv. in Phys.\/} {\bf 
62} 225

\bibitem{bfp_review} Majumdar S~N 2005 {\em Curr. Sci.\/} {\bf 89} 2076  

\bibitem{ADP2014}
Angelani L, Di Lionardo R, and Paoluzzi M 2014 {\em Euro. J. Phys. E\/}
{\bf 37}, 59

\bibitem{scacchi2017mean}
Scacchi A and Sharma A 2017 {\em arXiv preprint arXiv:1708.05591\/}


\bibitem{masoliver1992solutions}
Masoliver J, Porra J~M and Weiss G~H 1992 {\em Physical Review A\/} {\bf 45}
  2222

\bibitem{orsingher}
Orsingher E 1995, Random Oper. and Stoch. Equ., {\bf 3}, 9

\end{thebibliography}
\end{document}